\documentclass[aps,prb,onecolumn,eqsecnum]{revtex4-2}

\usepackage{graphicx}

\usepackage{bm}
\usepackage{amsmath}
\usepackage{amssymb}

\newcommand{\bvm}{\begin{verbatim}}
\newcommand{\evm}{\end{verbatim}}

\newcommand{\ocite}{\onlinecite}

\newcommand{\iy}{\infty}

\newcommand{\x}{\text}
\newcommand{\pd}{\partial}

\newcommand{\dg}{\dagger}
\newcommand{\lan}{\langle}
\newcommand{\ran}{\rangle}
\newcommand{\lt}{\left}
\newcommand{\rt}{\right}

\newcommand{\f}{\frac}
\newcommand{\tf}{\tfrac}

\newcommand{\sq}{\sqrt}
\newcommand{\lbl}{\label}

\newcommand{\cd}{\cdot}

\newcommand{\nm}{\hat{0}}
\newcommand{\um}{\hat{1}}

\newcommand{\Imx}{\text{Im}\,}

\newcommand{\tm}{\times}

\newcommand{\eq}[1]{Eq.~(\ref{eq:#1})}
\newcommand{\eqs}[2]{Eqs.~(\ref{eq:#1}) and (\ref{eq:#2})}
\newcommand{\eqss}[3]{Eqs.~(\ref{eq:#1}), (\ref{eq:#2}), and (\ref{eq:#3})}

\newcommand{\eqsd}[2]{Eqs.~(\ref{eq:#1})-(\ref{eq:#2})}
\newcommand{\eqn}[1]{(\ref{eq:#1})}
\newcommand{\eqsn}[2]{(\ref{eq:#1}) and (\ref{eq:#2})}

\newcommand{\secr}[1]{Sec.~\ref{sec:#1}}
\newcommand{\secsr}[2]{Secs.~\ref{sec:#1} and \ref{sec:#2}}

\newcommand{\figr}[1]{Fig.~\ref{fig:#1}}

\newcommand{\appr}[1]{Appendix~\ref{app:#1}}

\newcommand{\spc}{\mbox{ }}

\newcommand{\beq}{\begin{equation}}
\newcommand{\eeq}{\end{equation}}
\newcommand{\beqar}{\begin{eqnarray}}
\newcommand{\eeqar}{\end{eqnarray}}
\newcommand{\beqarn}{\begin{eqnarray*}}
\newcommand{\eeqarn}{\end{eqnarray*}}
\newcommand{\ba}{\begin{array}}
\newcommand{\ea}{\end{array}}
\newcommand{\bwt}{\begin{widetext}}
\newcommand{\ewt}{\end{widetext}}

\newcommand{\rarr}{\rightarrow}

\newcommand{\dx}{{\text d}}
\newcommand{\ex}{{\text e}}

\newcommand{\ix}{{\text i}}

\newcommand{\Ux}{{\text U}}

\newcommand{\ch}{\hat{c}}

\newcommand{\hh}{\hat{h}}

\newcommand{\ph}{\hat{p}}

\newcommand{\sh}{\hat{s}}

\newcommand{\Ch}{\hat{C}}

\newcommand{\Hh}{\hat{H}}

\newcommand{\Jh}{\hat{J}}

\newcommand{\Mh}{\hat{M}}

\newcommand{\Sh}{\hat{S}}
\newcommand{\Th}{\hat{T}}
\newcommand{\Uh}{\hat{U}}
\newcommand{\Vh}{\hat{V}}

\newcommand{\Xh}{\hat{X}}

\newcommand{\psih}{\hat{\psi}}
\newcommand{\tauh}{\hat{\tau}}
\newcommand{\chih}{\hat{\chi}}

\newcommand{\Psih}{\hat{\Psi}}

\newcommand{\prh}{\hat{\bar{p}}}

\newcommand{\Psir}{\bar{\Psi}}

\newcommand{\psirh}{\hat{\bar{\psi}}}
\newcommand{\Psirh}{\hat{\bar{\Psi}}}

\newcommand{\Cc}{\mathcal{C}}

\newcommand{\Ec}{\mathcal{E}}

\newcommand{\Hc}{\mathcal{H}}

\newcommand{\Jc}{\mathcal{J}}

\newcommand{\Nc}{\mathcal{N}}
\newcommand{\Oc}{\mathcal{O}}

\newcommand{\Tc}{\mathcal{T}}
\newcommand{\Uc}{\mathcal{U}}

\newcommand{\Wc}{\mathcal{W}}

\newcommand{\Hch}{\hat{\Hc}}

\newcommand{\Jch}{\hat{\Jc}}

\newcommand{\al}{\alpha}

\newcommand{\Ga}{\Gamma}
\newcommand{\de}{\delta}

\newcommand{\La}{\Lambda}

\newcommand{\ka}{\varkappa}

\newcommand{\eps}{\varepsilon}
\newcommand{\e}{\epsilon}

\begin{document}
\title{Formalism of general boundary conditions for continuum models}
\author{Maxim Kharitonov$^{1,2}$}
\address{$^1$Institute for Theoretical Physics and Astrophysics, University of W\"urzburg, 97074 W\"urzburg, Germany\\
$^2$Donostia International Physics Center (DIPC), Manuel de Lardizabal 4, E-20018 San Sebastian, Spain}

\begin{abstract}

Continuum models are particularly appealing for theoretical studies of bound states, due to simplicity of their bulk Hamiltonians. The main challenge on this path is a systematic description of the boundary, which comes down to determining proper boundary conditions (BCs). BCs are a consequence of the fundamental principle of quantum mechanics: norm conservation of the wave function, which leads to the conservation of the probability current at the boundary. The notion of {\em general BCs} arises, as a family of all possible BCs that satisfy the current-conservation principle. Ahari, Ortiz, and Seradjeh formulated a systematic derivation procedure of the general BCs from the current-conservation principle for the 1D Hamiltonian of the most general form. The procedure is based on the diagonalization of the current and leads to the universal ``standardized'' form of the general BCs, parameterized in a nonredundant one-to-one way by unitary matrices. In this work, we substantiate, elucidate, and expand this {\em formalism of general boundary conditions for continuum models}, addressing in detail a number of important physical and mathematical points. We provide a detailed derivation of the general BCs from the current-conservation principle and establish the conditions for when they are admissible in the sense that they describe a well-defined boundary, which is directly related to a subtle but crucial distinction between self-adjoint (hermitian) and only symmetric operators. We provide a natural physical interpretation of the structure of the general BCs as a scattering process and an essential mathematical justification that the formalism is well-defined for Hamiltonians of momentum order higher than linear. We discuss the physical meaning of the general BCs and outline the application schemes of the formalism, in particular, for the study of bound states in topological systems.

\end{abstract}
\maketitle

\section{Introduction\lbl{sec:intro}}

Bound states at boundaries, interfaces, and defects of bulk crystalline electron systems
have by now been firmly established as a widespread feature in a multitude of real materials
and theoretical models~\cite{Hasan,Alicea,Leijnse,Beenakker,Chiu2016,Armitage2017}
(we will use the term ``bound states'' for bound, edge, or surface states in a generalized sense for any dimension,
when system dimension is not explicitly specified or does not matter).
When the bulk has nontrivial topology, certain bound states are guaranteed, as per the concept of bulk-boundary correspondence~\cite{Chiu2016}.
Moreover, even systems that are topologically trivial in a rigorous sense
have been theoretically shown to exhibit robust bound states~\cite{KharitonovLSM,KharitonovQAH,KharitonovFGCM}.

{\em Continuum models} (CMs), whose bulk Hamiltonians are polynomials in momentum,
are particularly appealing for theoretical studies of bound states.
Most commonly, a CM  is intended as a simpler version of
a more complicated (``more'') microscopic system that arises in the low-energy limit of the latter and still captures its essential properties.
For a system with discrete translation symmetry, such as a crystal,
the Hamiltonian of the CM typically describes the low-energy expansion
of the underlying microscopic Hamiltonian about certain special points of interest in the Brillouin zone.
The multicomponent wave function of the CM consists of the envelopes of the quantum states at those points.
For nodal semimetals or superconductors, these special points of interest can be the nodes;
for topological insulators or nodeless superconductors with a generally gapped spectrum,
these can be points of gap closing at the topological phase transition.
CMs are thus particularly well-suited for the study of the vicinity of a topological phase transition, which is often also the most interesting regime.
For high enough spatial symmetry of the system, these special points of interest often happen to be the high-symmetry points in the Brillouin zone.

The main appeal of CMs is the relative analytical simplicity of their bulk Hamiltonians.
For a CM, one may take into account just enough ``degrees of freedom'',
i.e., wave-function components and momentum powers in the Hamiltonian (see \secr{H} for more details),
that capture the desired physical properties.
Often the lowest-order, linear-in-momentum Hamiltonians are sufficient to capture the behavior of interest in the bulk;
in particular, such Hamiltonians were studied earlier~\cite{Ryu2010,Chiu2016} as representative models for various topological symmetry classes.

\begin{figure}
\includegraphics[width=.44\textwidth]{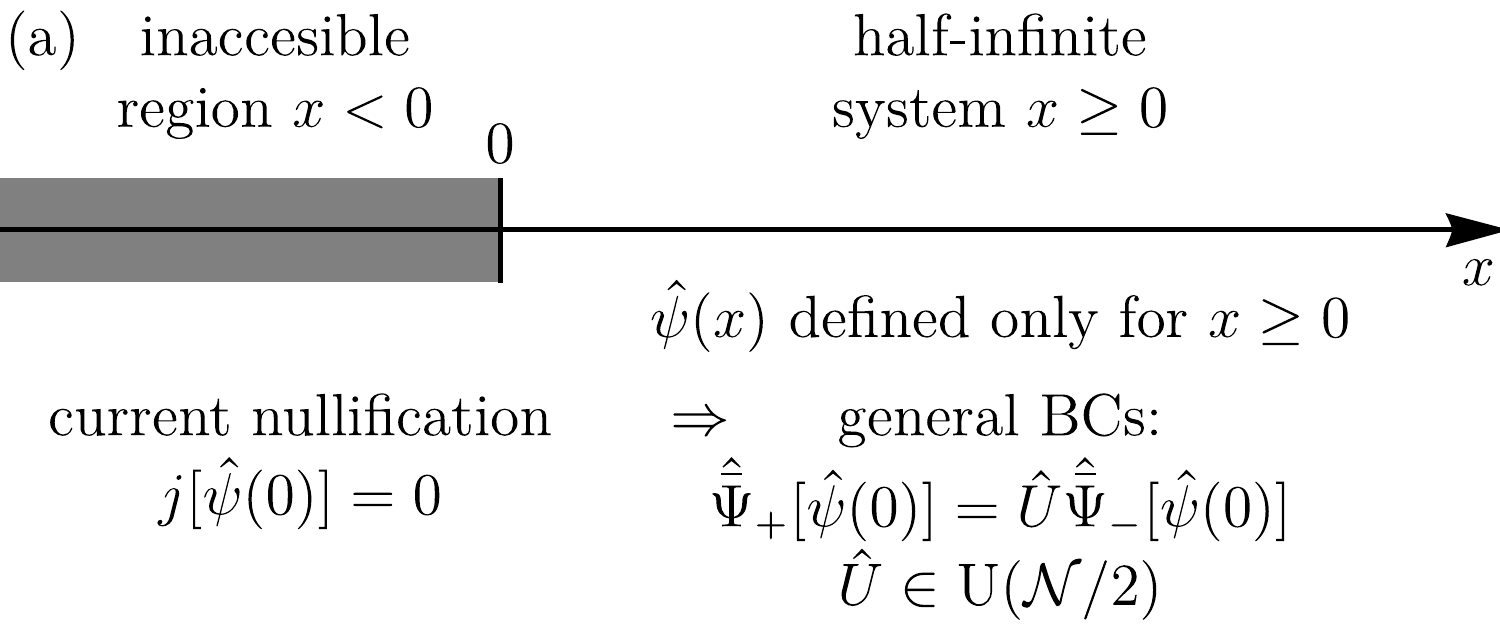}\\
\includegraphics[width=.18\textwidth]{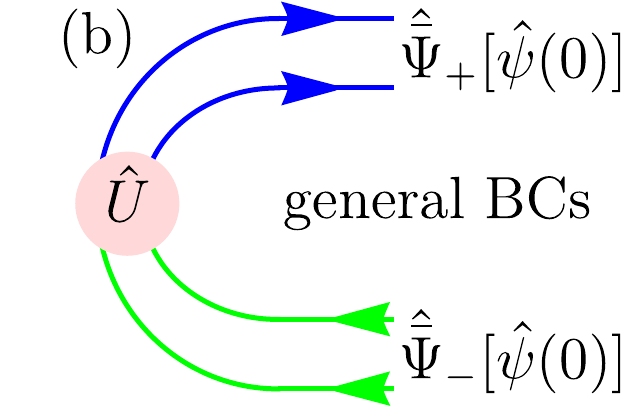}
\caption{
(a) The considered 1D half-infinite system occupying the region $x\geq0$
and (b) the physical interpretation of the general boundary conditions (BCs) \eqn{bc}
for the Hamiltonian $\Hh(\ph)$ [\eq{H}] of the most general form, polynomial in momentum up to some order $N$,
for the multicomponent wave function $\psih(x)$.
The wave-function components and their derivatives $\ph^n\psi_m(x)$ with $m=1,\ldots,M$ and $n=0,\ldots,N_m$
are arranged into the vectors $\Psirh_+[\psih(x)]$ and $\Psirh_-[\psih(x)]$ [\eq{Psirh}],
in terms of the $\Nc_\pm$ components of which the probability current $j[\psih(x)]$ is diagonalized and normalized
[\eq{jdn}] with the eigenvalues $\pm1$, respectively.
These components can be interpreted as chiral right- and left-moving ``waves'' (\secr{interpretation}),
which in this geometry are reflected from and incident upon the boundary, respectively.
The general BCs \eqn{bc}, well-defined only in the case of equal numbers $\Nc_+=\Nc_-=\Nc/2$,
is the matrix relation between the vectors $\Psirh_+[\psih(0)]$ and $\Psirh_-[\psih(0)]$ at the boundary via a unitary matrix $\Uh\in\Ux(\Nc/2)$.
This form has a natural physical interpretation as a scattering process between the chiral right- and left-moving waves.
The family of general BCs is parameterized in a nonredundant one-to-one way by all possible unitary matrices $\Uh$.
}
\lbl{fig:system}
\end{figure}

Since CMs can provide a tremendous simplification of the underlying microscopic system,
it would be desirable to exploit this advantage not only for the analysis of the bulk properties, but also for the study of bound states.
However, despite this appeal, application of CMs to the study of bound states has generally been rather underdeveloped and non-systematic,
the main reason being the challenge of describing the sample boundary within CMs.
In fact, it is quite common, especially in the studies of topological systems, that the analysis of the bulk properties is performed within a CM,
but the bound states are studied via some ``regularization'' procedure, e.g., using instead a lattice model,
whose Hamiltonian reduces to that of the CM in the low-energy limit.
The reason is that for lattice models the termination of the sample can easily be described as the absence of sites.
However, such an approach has the following drawbacks.
(i) The analysis of bound states is usually done numerically for a large finite-size system (where finite-size effects impose additional limitations).
(ii) Typically, some specific choice of the lattice termination is made.
If one resorts to just one or a few possibilities, some possible bound-state structures could be overlooked.
On the other hand, considering different lattice terminations can be computationally costly.
(iii) Sometimes, the lattice model introduced in the regularization procedure is different (in structure and sometimes even in symmetry)
from the true original lattice of the system of interest (because the former is still simpler than the latter).
This way, in the end, the bound states are actually studied not within a CM, but within a {\em different} model.
Such approach to calculating bound states therefore partially defeats the point of considering the CM in the first place,
since the main advantage of the latter as being a technically much simpler model of the system is lost.

Meanwhile, a description of the boundary fully within CMs is entirely possible: the problem reduces to deriving proper {\em boundary conditions} (BCs).
It has been understood for quite some
time~\cite{AkhiezerGlazman,ReedSimon,Berry,BerezinShubin,Bonneau,Tokatly,McCann,AkhmerovPRL,AkhmerovPRB,Ostaay,Hashimoto2016,Hashimoto2019,Ahari,
KharitonovLSM,KharitonovQAH,Seradjeh,Walter,Shtanko,KharitonovSC,Enaldiev2015,Volkov2016,Devizorova2017,KharitonovFGCM}
that BCs arise as a consequence of the fundamental principle of quantum mechanics:
conservation of the wave-function norm upon the time evolution described by the Schr\"odinger equation.
For a system with a boundary, this imposes constraints not only on the form of the Hamiltonian operator,
but also on the Hilbert space of allowed wave functions.
The latter constraint has the form of the nullification of the probability current at the boundary.
This constraint can be resolved in the form of
linear homogeneous relations between the wave-function components and their derivatives at the boundary,
and these relations are commonly referred to as the {\em boundary conditions} (BCs).
This understanding naturally leads to the notion of {\em general BCs},
as a family of all possible BCs (for a given Hamiltonian) that resolve the current-nullification constraint.

Previously, general BCs have been derived from the current-conservation principle
and used to study the bound-, edge-, or surface-state structures for specific continuum models:
for the 1D quadratic-in-momentum one-component model~\cite{Bonneau} (a textbook nonrelativistic Schr\"odinger particle);
for the 3D quadratic-in-momentum  multi-component model~\cite{Tokatly}, in the context of semiconductors;
for the linear-in-momentum two-component model,
in the context of 2D semimetals~\cite{Berry,KharitonovLSM,KharitonovFGCM}, 2D quantum anomalous Hall (QAH) systems~\cite{KharitonovQAH},
graphene~\cite{Walter}, 1D insulators~\cite{Ahari,KharitonovFGCM}, and 3D Weyl semimetals~\cite{Hashimoto2016,Hashimoto2019,KharitonovFGCM};
for the linear-in-momentum four-component model of graphene~\cite{McCann,AkhmerovPRL,AkhmerovPRB,Ostaay}
and 3D Dirac materials~\cite{Shtanko};
for the 3D quadratic-in-momentum two-component model of Weyl semimetals~\cite{Seradjeh};
for the 1D linear-in-momentum four-component model of superconductors with one Fermi surface~\cite{KharitonovSC}.
Also, general BCs with some symmetry constraints have been derived for quadratic- and linear-in-momentum four-component models
of 2D and 3D topological insulators and Dirac materials~\cite{Enaldiev2015,Volkov2016} and two-node 3D Weyl semimetals~\cite{Devizorova2017}.

Though the general understanding of the origin of BCs for CMs has existed for some time,
the approaches of the above works (the procedure of finding BCs, the form and parametrization of BCs)
have been tailored to specific, often simplest Hamiltonians. 
On the other hand, in Ref.~\ocite{Ahari}, Ahari, Ortiz, and Seradjeh formulated a systematic derivation procedure
of the general BCs from the current-conservation principle for the 1D continuum model with the translation-symmetric Hamiltonian of the most general form,
with any number of wave-function components and any order of momentum.
The central technical advancement was to present the probability current in the universal diagonal and normalized form.
This leads to the ``standardized'' universal form of the family of the general BCs,
which are parameterized in a nonredundant one-to-one way by unitary matrices.

\section{Structure of the paper and summary of the results\lbl{sec:results}}

In this work, we substantiate, elucidate, and expand
this {\em formalism of general boundary conditions (BCs) for continuum models (CMs)}, initiated in Ref.~\ocite{Ahari},
addressing in detail and clarifying a number of important physical and mathematical points.
The paper is organized as follows.

In \secr{model}, we present the Hamiltonian for the 1D translation-symmetric CM of the most general form
and the half-infinite system, for which the general BCs will be derived.

In \secr{herm}, we demonstrate how the current-conservation principle arises from the fundamental
principle of quantum mechanics, the norm conservation of the wave function.

\secr{formalism} is devoted to deriving the general BCs and providing the necessary justification.
In \secr{j}, the main properties of the probability current are presented.
In \secr{jdiagonalization}, we reproduce the key initial step of the derivation
that brings the current to the universal diagonal and normalized form.
In \secr{derivation}, we derive the families of general BCs that nullify the current at the boundary.
In \secsr{twoboundaries}{halfinfinite}, we demonstrate that BCs
nullifying the current are still not always admissible in the sense that they represent a well-defined boundary
and establish the corresponding conditions.
In \secr{symmvsherm}, we demonstrate that this distinction between which BCs that nullify the current
are admissible or not coincides with the distinction between the Hilbert spaces over which the Hamiltonian is self-adjoint (hermitian) or only symmetric.

In \secr{interpretation}, we provide a natural physical interpretation of the structure of the general BCs
as a scattering process at the boundary between the incident (left-moving) and reflected (right-moving) modes of the wave function,
illustrated in \figr{system}(b).

In \secr{justification},
we study in detail the structure of the current matrix
and provide an essential mathematical justification that the formalism is well-defined for Hamiltonians of momentum order higher than linear;
this inevitably requires introducing a fictitious length scale.

In \secr{method}, we present the general, analytical or semi-analytical, method of calculating the bound states
that becomes available within the framework of CMs with BCs.

In \secr{realization}, we explain the physical meaning of the general BCs in terms of their relation to the underlying microscopic models
and outline the related systematic low-energy-expansion procedure of deriving low-energy continuum models with BCs from microscopic models.

In \secr{genbsstr}, we explain that the general BCs lead to general bound-state structures
and discuss other related advantages that the formalism provides for the study of bound states.

In \secr{schemes}, we present two important application schemes of the formalism to {\em families} of bulk systems.
In the first scheme, the Hamiltonian is of the most general form satisfying some set of symmetries
and the BCs are of the most general form, satisfying only the current-conservation principle.
In the second scheme (already formulated recently in Ref.~\cite{KharitonovFGCM}),
both the Hamiltonian and BCs are of the most general form satisfying the same set of symmetries.
The second scheme is perfectly suited for the study of bound states in topological systems.

In \secsr{N2M1}{N1M2}, we illustrate the application of the formalism to the two models
with the minimal number of degrees of freedom necessary to have a well-defined boundary:
one-component quadratic-in-momentum Hamiltonian and two-component linear-in-momentum Hamiltonian.
In both models, a bound state is present in a half of the $\Ux(1)$ parameter space of the general BC.
For the quadratic model, we present two examples of possible realization of the general BC by potentials.
Using these two examples, in \secr{connection}, we demonstrate the relation in the structure of the general BCs
between the number of wave-function components and the order of momentum in the Hamiltonian.

In \secr{generalizations}, we discuss possible generalizations and extensions of the formalism of general BCs
to higher dimensions and hybrid systems with junctions and interfaces.

Concluding remarks are presented in \secr{conclusion}.

\section{General 1D continuum model \lbl{sec:model}}

\subsection{Hamiltonian \lbl{sec:H}}

This work is devoted to the formalism of general BCs specifically in one dimension (1D);
generalization to higher dimensions should be possible, as briefly discussed in \secr{generalizations}.

In 1D, the most general form of the single-particle Hamiltonian of a CM with a translation-symmetric bulk is a polynomial matrix function
\beq
	\Hh(\ph)=\sum_{n=0}^N \hh_n \ph^n
\lbl{eq:H}
\eeq
of the momentum operator
\beq
	\ph=-\ix\pd_x
\lbl{eq:p}
\eeq
(we use the units in which the Planck constant $\hbar=1$)
for a multicomponent wave function
\beq
	\psih(x)=\lt(\ba{c} \psi_1(x)\\\psi_2(x)\\\ldots\\\psi_{M-1}(x)\\\psi_M(x)\ea\rt).
\lbl{eq:psi}
\eeq
Throughout, $M$ will denote an arbitrary number of wave-function components $\psi_m(x)$, $m=1,\dots,M$; hence, $\hh_n$ are $M\tm M$ matrices.
The top order of momentum $\ph$ in $\Hh(\ph)$ for the component $\psi_m(x)$ will be denoted as $N_m\geq 1$;
in general, $N_m$ could differ for different $m$.
The top order of momentum in $\Hh(\ph)$ among all components is the maximum $N=\max_m N_m$.

Although the (real) momentum eigenvalues $p$ are formally unrestricted in a CM,
it should be kept in mind that, if such CM represents a crystal with only a discrete translation symmetry,
the CM is valid only for small enough momenta and respective small enough energies.
In particular, for a gapped system, the gaps between the bands must be much smaller than the full widths of the bands.
Within this validity range, the CM is a rigorous asymptotic limit of the underlying microscopic model.
The momentum operator $\ph$ corresponds to the deviation of quasimomentum from the expansion points of interest of the latter.

\subsection{Half-infinite system, inaccessible region \lbl{sec:system}}

We consider a half-infinite system occupying the region $x\geq 0$, \figr{system}.
The wave function $\psih(x)$ [\eq{psi}] is defined only in this region and is undefined in $x<0$. Physically, this means the following.
As already mentioned above, typically, a continuum model arises as a low-energy limit of some (``more'') microscopic model.
The wave function of this underlying microscopic model (which is at least implied, even when not specified) is defined for all $x$.
In the region $x\geq0$, this model has low-energy excitations, which are described  by the continuum model of interest,
with the wave function $\psih(x)$ [\eq{psi}] and the Hamiltonian $\Hh(\ph)$ [\eq{H}].
Whereas in the region $x<0$, the microscopic model has a gap~\cite{gapcomment}
in the spectrum that is much larger than the relevant low-energy scale.
As a result, the microscopic wave function decays rapidly into the region $x<0$.
The half-space $x<0$ is therefore the region ``inaccessible'' for the low-energy excitations that the wave function $\psih(x)$ describes.

\section{Conservation of wave-function norm as the origin of boundary conditions \lbl{sec:herm}}

In this section, we demonstrate how BCs arise as a consequence of the fundamental principle of quantum mechanics: conservation of the wave-function norm.

As the starting point of our analysis, we assume no constraints on the form of the Hamiltonian operator $\Hh(\ph)$ [\eq{H}],
i.e., that $\hh_n$ therein are initially arbitrary complex matrices.
Also, no initial boundary constraints on the wave functions $\psih(x)$ are assumed:
all wave functions initially belong to the larger ``embedding'' Hilbert space $\Upsilon_0$ of normalizable wave functions
with appropriate smoothness properties for a given $\Hh(\ph)$.
All necessary restrictions on both the Hamiltonian and Hilbert space should and will follow from the norm-conservation principle.

A fundamental principle of the theory of quantum mechanics is that the norm
\beq	
	\lan \psih(t),\psih(t)\ran=\int_0^{+\iy}\dx x\, \psih^\dg(x,t)\psih(x,t)
\lbl{eq:norm}
\eeq
of the time-dependent wave function $\psih(t)=\psih(x,t)$ is conserved upon time evolution described by the Schr\"odinger equation
\beq
	\ix\pd_t\psih(x,t)=\Hh(\ph)\psih(x,t).
\lbl{eq:scheq}
\eeq
In other words, the norm \eqn{norm} is a time-independent constant, which is equivalent to its time derivative being zero:
\beq
	\pd_t\lan\psih(t),\psih(t)\ran=0.
\lbl{eq:dtnorm=0}
\eeq
Here,
\[
	\lan \psih_a,\psih_b\ran=\int_0^{+\iy}\dx x\, \psih_a^\dg(x)\psih_b(x)
\]
is the scalar product in the embedding Hilbert space $\Upsilon_0$,
given by the integral over the region $x\geq 0$ of the half-infinite system, described in \secr{H}.
The norm and the scalar product are well-defined only for normalizable wave functions, that decay $\psih(x)\rarr\nm$ at $x\rarr +\iy$
and in such a way that the integral converges (in reality, only the ``wave packets'' consisting of normalizable wave functions are ever realized).
Throughout, $\dg$ denotes hermitian conjugation (complex conjugation $^*$ and transposition)
of a matrix of any size, including the column vector \eqn{psi} of the wave function;
so, $\psih^\dg(x)=(\psi_1^*(x),\ldots,\psi_M^*(x))$ is a row vector and $\ph^\dg=\ph^*=-\ph$ [\eq{p}] below.

Using the Schr\"odinger equation, one obtains that this is equivalent to the following condition being satisfied:
\beq
	\ix\pd_t\lan\psih(t),\psih(t)\ran=\lan\psih(t),\Hh\psih(t)\ran-\lan\Hh\psih(t),\psih(t)\ran=0.
\lbl{eq:dt=0}
\eeq
For this to be satisfied for any time-dependent solution $\psih(x,t)$, the condition
\beq
	\lan\psih,\Hh\psih \ran=\lan\Hh\psih,\psih\ran
\lbl{eq:Hsymmdef}
\eeq
has to be satisfied for any wave function $\psih(x)\in\Upsilon$ in the sought Hilbert space $\Upsilon\subset\Upsilon_0$.
The latter is the definition of a {\em symmetric} operator $\Hh(\ph)$ over the Hilbert space $\Upsilon$;
we forewarn that this is {\em not} the definition of a {\em hermitian} operator, the term we use synonymously to {\em self-adjoint};
this important distinction will be discussed in \secr{symmvsherm}.
We note that, importantly, \eq{Hsymmdef} is a constraint on {\em both} the operator $\Hh(\ph)$ and the Hilbert space $\Upsilon$.

Consider the time evolution of the more general quantity than the norm \eqn{norm}
(this is necessary for dealing with the hermitian Hamiltonian operator in \secr{symmvsherm}): the local-in-coordinate sesquilinear form
\[
	\rho[\psih_a(x,t),\psih_b(x,t)]=\psih_a^\dg(x,t)\psih_b(x,t)
\]
of two different wave functions $\psih_{a,b}(x,t)$, both satisfying the Schr\"odinger equation \eqn{scheq}.
Using the Schr\"odinger equation, we obtain (temporarily dropping the arguments for brevity)
\beq
	\ix\pd_t(\psih_a^\dg\psih_b)=\sum_{n=0}^N\lt[\psih_a^\dg\hh_n\ph^n\psih_b-(\ph^n\psih_a)^\dg\hh_n^\dg\psih_b\rt]
	=\sum_{n=0}^N\lt\{\psih_a^\dg\hh_n'\ph^n\psih_b-(\ph^n\psih_a)^\dg\hh_n'\psih_b
	- \ix[\psih_a^\dg\hh_n''\ph^n\psih_b+(\ph^n\psih_a)^\dg\hh_n''\psih_b]\rt\},
\lbl{eq:dtab}
\eeq
where the initially arbitrary complex matrices
\beq
	\hh_n=\hh_n'-\ix\hh_n''
\lbl{eq:hninit}
\eeq
are presented in terms of hermitian matrices: $\hh_n'^\dg=\hh_n'$, $\hh_n''^\dg=\hh_n''$.
We recognize that the first part on the right-hand side of \eq{dtab} is presentable as the full coordinate derivative
of the quantity
\beq
	j[\psih_a(x,t),\psih_b(x,t)]=\sum_{n=1}^N j_n[\psih_a(x,t),\psih_b(x,t)],\spc
	j_n[\psih_a(x,t),\psih_b(x,t)]=\sum_{n'=0}^{n-1} (\ph^{n'}\psih_a(x,t))^\dg \hh_n' (\ph^{n-1-n'}\psih_b(x,t)),
\lbl{eq:jabinit}
\eeq
so that \eq{dtab} can be presented as
\beq
	\pd_t \rho[\psih_a(x,t),\psih_b(x,t)]=-\pd_x j[\psih_a(x,t),\psih_b(x,t)]+q[\psih_a(x,t),\psih_b(x,t)]
\lbl{eq:dtrhoab}
\eeq
with
\[
	q[\psih_a(x,t),\psih_b(x,t)]=-\sum_{n=0}^N\lt[\psih_a^\dg(x,t)\hh_n''\ph^n\psih_b(x,t)+(\ph^n\psih_a(x,t))^\dg\hh_n''\psih_b(x,t)\rt].
\]
In particular, for the same wave function $\psih_a(x,t)=\psih_b(x,t)=\psih(x,t)$, this form
\[
	\rho[\psih(x,t)]=\psih^\dg(x,t)\psih(x,t)
\]
is the probability density and \eq{dtrhoab} takes the form of the continuity equation
\beq
	\pd_t \rho[\psih(x,t)]=-\pd_x j[\psih(x,t)]+q[\psih(x,t)]
\lbl{eq:dtrho}
\eeq
with the real probability current
\beq
	j[\psih(x,t)]=\sum_{n=1}^N j_n[\psih(x,t)], \spc j_n[\psih(x,t)]=\sum_{n'=0}^{n-1} (\ph^{n'}\psih(x,t))^\dg \hh_n' (\ph^{n-1-n'}\psih(x,t))
\lbl{eq:jinit}
\eeq
and source
\beq
	q[\psih(x,t)]=-\sum_{n=0}^N[\psih^\dg(x,t)\hh_n''\ph^n\psih(x,t)+(\ph^n\psih(x,t))^\dg\hh_n''\psih(x,t)]
\lbl{eq:q}
\eeq
terms.

Using \eqs{dt=0}{dtrho}, for the deviation from the equality \eqn{Hsymmdef}, we have
\beq
	\lan\psih,\Hh\psih \ran-\lan\Hh\psih,\psih\ran
	=\ix\int_0^{+\iy}\dx x\, q[\psih(x)]+\ix j[\psih(x=0)],
\lbl{eq:d=qj}
\eeq
where the current $j[\psih(x=+\iy)]=0$ at infinity vanishes for the normalizable wave function.

There are two contributions on the right-hand side of \eq{d=qj}, which must vanish individually.
The first one is a {\em bulk} contribution: an integral of the source term \eqn{q}.
For the source term $q[\psih(x)]\equiv0$ to vanish identically for any wave function, the matrices
\[
	\hh_n''=\nm
\]
have to vanish, which means that the matrices $\hh_n=\hh_n'$ at each power $\ph^n$ of momentum in the Hamiltonian must be hermitian [\eq{hninit}]:
\beq
	\hh_n^\dg=\hh_n, \spc n=0,\ldots,N.
\lbl{eq:hnherm}
\eeq
As expected, this is equivalent to the Hamiltonian $\Hh(p)$ being a hermitian matrix when momentum $p$ is a real number:
\beq
	\Hh^\dg(p)=\Hh(p).
\lbl{eq:Hherm}
\eeq
Nullification of the bulk contribution to the time derivative \eqn{dt=0} of the norm
therefore provides a constraint on the form of the Hamiltonian as a local differential operator.
This standard requirement is assumed satisfied in the rest of the paper.

For a system with a boundary, this is, however, only a necessary, but not a sufficient condition for the wave-function norm to be conserved.
The second contribution to \eq{d=qj} is a boundary contribution given by the flow of the probability current.
For the time derivative of the norm of any solution to the Schr\"odinger equation to vanish [\eq{dtnorm=0}],
the current through the boundary also has to vanish for every wave function $\psih(x)$ in the sought Hilbert space(s) $\Upsilon$:
\beq
	j[\psih(x=0)]=0.
\lbl{eq:j=0}
\eeq
We see that while the vanishing of the bulk contribution to the time derivative \eqn{dt=0} [via \eq{d=qj}]
is a restriction on the form of the Hamiltonian as a local differential operator,
the vanishing of the boundary contribution is ultimately a restriction on the Hilbert space $\Upsilon \subset \Upsilon_0$.

Since the current $j[\psih(x=0)]$ involves only the wave-function components and their derivatives at the boundary [\eq{jinit}, see also \secr{j}],
such Hilbert spaces $\Upsilon$ can be specified in the form of linear homogeneous relations between these quantities
that resolve the current nullification constraint \eqn{j=0}.
And these relations are commonly referred to as the {\em boundary conditions} (BCs).
Therefore, BCs are essentially a way of specifying Hilbert spaces over which the current at the boundary is nullified.
According to \eqs{Hsymmdef}{d=qj}, these are the Hilbert spaces over which the Hamiltonian is symmetric.

The first subsections of \secr{formalism} are devoted to finding all general BCs, i.e.,
all families of all possible BCs that resolve the current-nullification constraint \eqn{j=0}.
Next, we observe that the situation turns out to be more subtle,
as not any BCs nullifying the current are {\em``admissible''} and deliver an {\em``admissible''} Hilbert space,
in the sense that they represent a system with a well-defined boundary.
For some systems, there are no admissible BCs at all, which means that for such 1D systems a boundary cannot be introduced.
Which BCs and the Hilbert spaces they specify should be deemed admissible has itself to be argued.
We demonstrate that these nuances are directly related to the subtle but crucial distinction
between {\em self-adjoint} ({\em hermitian}) and {\em only symmetric} Hamiltonian operators.
Establishing these conditions and formulating appropriate arguments is part of the development of the formalism, presented in this work.

\section{Formalism of general boundary conditions \lbl{sec:formalism}}
\subsection{Probability current \lbl{sec:j}}

Here, we summarize the main properties of the probability current.
Under the assumed standard constraints \eqsn{hnherm}{Hherm} on the form of the Hamiltonian operator,
the quadratic form of the probability current for one wave function $\psih(x)$ for the Hamiltonian $\Hh(\ph)$ [\eq{H}] reads
\beq
	j[\psih(x)]=\sum_{n=1}^N j_n[\psih(x)],
\lbl{eq:j}
\eeq
\beq
	j_n[\psih(x)]=\sum_{n'=0}^{n-1} (\ph^{n'}\psih(x))^\dg \hh_n (\ph^{n-1-n'}\psih(x)).
\lbl{eq:jn}
\eeq
Here, $j_n[\psih(x)]$ is the respective contribution from the term $\hh_n\ph^n$ in the Hamiltonian; in particular,
\beq
	j_1[\psih]=\psih^\dg\hh_1\psih,
	\spc j_2[\psih]=\psih^\dg\hh_2(\ph\psih)+(\ph\psih)^\dg\hh_2\psih,
	\spc
	j_3[\psih]=\psih^\dg\hh_3(\ph^2\psih)+(\ph\psih)^\dg\hh_3(\ph\psih)+(\ph^2\psih)^\dg\hh_3\psih.
\lbl{eq:j123}
\eeq
The structure \eqn{jn} of $j_n[\psih(x)]$ can be readily understood:
for a plane-wave wave function $\psih(x)\propto \ex^{\ix px}$, when the momentum operator $\ph\rarr p$ becomes a real number,
the current is the derivative $\pd_p \Hh(p)=\sum_{n=1}^N \hh_n np^{n-1}$ of the Hamiltonian
and \eq{jn} is the properly symmetrized operator version of each contribution $\hh_n n p^{n-1}$.
Note that the zero-order term $\hh_0$ in the Hamiltonian does not contribute to the current.
The current \eqn{j} is a {\em local}-in-coordinate real quadratic form of wave-function components ($n=0$) and their derivatives ($n>0$)
\beq
	\ph^n\psi_m(x) \mbox{ with $m=1,\ldots,M$ and $n=0,\ldots,N_m-1$},
\lbl{eq:pnpsi}
\eeq
which should be treated as its independent variables.
Note that it will be technically more convenient {\em not} to spell out the momentum operator \eqn{p} in terms of the derivative $\pd_x$;
for brevity, we will still refer to these quantities as {\em derivatives}.
The total number of the wave-function components and their derivatives \eqn{pnpsi} entering the current \eqn{j} is
\beq
	\Nc=\sum_{m=1}^M N_m.
\lbl{eq:Nc}
\eeq
These quantities \eqn{pnpsi} can be viewed as and will be referred to as the {\em degrees of freedom} of the CM.
Their total number $\Nc$ is equal to the number of linearly independent particular solutions to
the stationary Schr\"odinger equation $\Hh(\ph)\psih(x)=\e\psih(x)$ at some energy $\e$.

\subsection{Diagonalization of the probability current \lbl{sec:jdiagonalization}}

The first technical goal of the formalism is finding all BCs that resolve the current-nullification constraint \eqn{j=0}.
The central technical advancement made in Ref.~\cite{Ahari} was to present the probability current [\eqs{j}{jn}] in the diagonal normalized form.
We reproduce this step here. The quadratic form of the probability current is first presented in the matrix form
\beq
	j[\psih(x)]=\Psih^\dg[\psih(x)]\Jh\Psih[\psih(x)],
\lbl{eq:Jdef}
\eeq
where the wave-function components and their derivatives \eqn{pnpsi}
are arranged into a vector $\Psih[\psih(x)]$ of size $\Nc$ [\eq{Nc}] and $\Jh$ is a hermitian $\Nc\tm\Nc$ matrix, $\Jh^\dg=\Jh$.
These quantities will be specified explicitly and explored in detail in \secr{justification}.
Importantly, as also discussed in \secr{justification}, for a well-defined bulk spectrum, $\Jh$ has to be nondegenerate.
The current matrix $\Jh$ can therefore be diagonalized in the form
\[
	\Jch=\Sh^\dg\Jh\Sh=\x{diag}(\Jc_{+,1},\ldots,\Jc_{+,\Nc_+},\Jc_{-,1},\ldots,\Jc_{-,\Nc_-})
\]
with $\Nc_+$ positive $\Jc_{+,\nu_+}>0$ ($\nu_+=1,\ldots,\Nc_+$) and $\Nc_-$ negative $\Jc_{-,\nu_-}<0$ ($\nu_-=1,\ldots,\Nc_-$) eigenvalues, such that
\beq
	\Nc_++\Nc_-=\Nc.
\lbl{eq:Nc+-}
\eeq
Here,
\beq
	\Sh=\lt(\sh_{+,1},\ldots,\sh_{+,\Nc_+},\sh_{-,1},\ldots,\sh_{-,\Nc_-}\rt)
\lbl{eq:S}
\eeq
is a unitary matrix whose columns $\sh_{\pm,\nu_\pm}$ are the normalized eigenvectors of $\Jh$
with positive and negative eigenvalues $\Jc_{\pm,\nu_\pm}$, respectively.
The current is then presented in the diagonal form
\beq
	j[\psih(x)]=\sum_{\nu_+=1}^{\Nc_+}\Jc_{+,\nu_+}\Psi_{+,\nu_+}^*[\psih(x)]\Psi_{+,\nu_+}[\psih(x)]
	+\sum_{\nu_-=1}^{\Nc_-} \Jc_{-,\nu_-}\Psi_{-,\nu_-}^*[\psih(x)]\Psi_{-,\nu_-}[\psih(x)],
\lbl{eq:jd}
\eeq
where
\beq
	\Psi_{\pm,\nu_\pm}[\psih(x)]=\sh_{\pm,\nu_\pm}^\dg\Psih[\psih(x)]
\lbl{eq:Psipm}
\eeq
are the projections of the vector $\Psih[\psih(x)]$ onto the eigenvectors $\sh_{\pm,\nu_\pm}$.

One can further perform the nonunitary ``stretch'' transformation (note that it also changes the physical dimension)
\beq
	\Psir_{\pm,\nu_\pm}[\psih(x)]=\sq{|\Jc_{\pm,\nu_\pm}|}\Psi_{\pm,\nu_\pm}[\psih(x)],
\lbl{eq:Psir}
\eeq
to bring the current to the quadratic ``normalized'' form
\beq
	j[\psih(x)]
	=\Psirh_+^\dg[\psih(x)]\Psirh_+[\psih(x)]-\Psirh_-^\dg[\psih(x)]\Psirh_-[\psih(x)]
\lbl{eq:jdn}
\eeq
with $\pm 1$ eigenvalues in terms of the rescaled projections $\Psir_{\pm,\nu_\pm}[\psih(x)]$, which we join into the vectors
\beq
	\Psirh_\pm[\psih(x)]=\lt(\ba{c} \Psirh_{\pm,1}[\psih(x)]\\\ldots\\\Psirh_{\pm,\Nc_\pm}[\psih(x)]\ea\rt)
\lbl{eq:Psirh}
\eeq
of sizes $\Nc_\pm$, respectively.

Note that according to the signs of their contributions $\pm\Psir_{\pm,\nu_\pm}^*[\psih(x)]\Psir_{\pm,\nu_\pm}[\psih(x)]\gtrless0$
to the current \eqn{jdn}, $\Psir_{\pm,\nu_\pm}[\psih(x)]$
can already now be interpreted as chiral ``waves'' (or ``modes'') propagating in the positive and negative $x$ direction, respectively,
i.e., as right- and left-moving waves (or, for brevity, simply, right- and left-movers), in the geometry of \figr{system}.
We adopt this terminology from now on and substantiate this interpretation more in \secr{interpretation}.

\subsection{Derivation of the general boundary conditions \lbl{sec:derivation}}

As the next step, we now find the families of all possible BCs that resolve the current nullification constraint \eqn{j=0}.
Presenting the current in the universal diagonal normalized form \eqn{jdn}, as performed in Ref.~\ocite{Ahari},
completely ``standardizes'' and unifies the problem:
instead of the initial degrees of freedom, the wave-function components and their derivatives $\ph^n\psi_m(0)$ [\eq{pnpsi}],
their linear combinations $\Psir_{\pm,\nu_\pm}[\psih(0)]$ [\eq{Psir}] now become the independent variables,
in terms of which the current is presented in the universal form \eqn{jdn},
which is fully specified for any Hamiltonian just by their numbers $\Nc_\pm$.
Consequently, finding the BCs in terms of $\Psir_{\pm,\nu_\pm}[\psih(0)]$ variables then solves the problem for {\em any} Hamiltonian.
The BCs can then be simply expressed in terms of the original degrees of freedom \eqn{pnpsi} via diagonalization formulas \eqsn{Psipm}{Psir}.

We first look for all largest subspaces (of the highest dimension) of the vector space of
\[
	\lt(\ba{c} \Psirh_+[\psih(0)] \\ \Psirh_-[\psih(0)] \ea\rt),
\]
over which the current form \eqn{jdn} is nullified identically.
Each such subspace can be specified by $\La$ linearly independent relations
for the components of $\Psirh_\pm[\psih(0)]$ (the to-be-derived correct form of which will become the BCs).
The  ``right'' number $\La$ of such relations (and what that means in the first place),
which determines the dimension $\Nc_++\Nc_--\La$ of these subspaces,
is itself to be determined and is a nontrivial question; arguments will follow.
The system of $\La$ relations of the most general form can be presented in the matrix form
\beq
	\Ch_+\Psirh_+[\psih(0)]=\Ch_-\Psirh_-[\psih(0)],
\lbl{eq:CPsi}
\eeq
where $\Ch_\pm$ are arbitrary matrices of dimensions $\La\tm\Nc_\pm$, respectively.

We first notice that the matrices $\Ch_\pm$ cannot have the ranks lower than $\Nc_\pm$, respectively.
Indeed, suppose $\Ch_+$ has a rank lower than $\Nc_+$. Then, taking $\Psirh_-[\psih(0)]=\nm$ as a null vector (of size $\Nc_-$),
the system \eqn{CPsi} reduces to $\Ch_+\Psirh_+[\psih(0)]=\nm$ (where $\nm$ is a null vector of size $\La$), which has a nonzero solution,
for which the current $j[\psih(0)]=\Psirh_+[\psih(0)]\Psirh_+[\psih(0)]$ [\eq{jdn}] is nonzero.
The proof for the minimal rank $\Nc_-$ of $\Ch_-$ is the same.
Hence, the number $\La$ of linearly independent relations must be equal or greater than the dimension of the larger vector:
\[
	\La\geq\max(\Nc_+,\Nc_-),
\]
while $\La<\max(\Nc_+,\Nc_-)$ is the situation of not enough BCs, when the current cannot be nullified identically.

Without loss of generality, we can and will assume $\Nc_+\geq\Nc_-$ from now on,
since $\Psirh_\pm[\psih(0)]$ enter the current form \eqn{jdn}
in a completely equivalent way, as far as its nullification is concerned.
Consider the case $\La=\Nc_+$, when the number of BCs is equal to the larger of $\Nc_\pm$.
In this case, $\Ch_+$ has to be a square nondegenerate matrix.
Then, the system of $\Nc_+$ relations of the most general form can be presented as
\beq
	\Psirh_+[\psih(0)]=\Vh\Psirh_-[\psih(0)],
\lbl{eq:VPsi}
\eeq
with an arbitrary $\Nc_+\tm\Nc_-$ matrix $\Vh$ (related to the previous matrices as $\Vh=\Ch_+^{-1}\Ch_-$, which are, however, no longer needed).
This convention also removes the unwanted redundancy of the BCs parametrization due to equivalence transformations present in \eq{CPsi},
since these do not change the subspace that the relations define.

Substituting the form \eqn{VPsi} into the expression \eqn{jdn} for the current, we obtain the form
\[
	j[\psih(0)] =\Psirh_-^\dg[\psih(0)](\Vh^\dg\Vh-\um)\Psirh_-[\psih(0)]
\]
with only $\Psih_-[\psih(0)]$ left, which have now become the only remaining independent variables.
Here, $\um$ is an $\Nc_-\tm\Nc_-$ unit matrix.
This form has to vanish [\eq{j=0}] for every $\Psirh_-[\psih(0)]$, i.e.,
it must be a null form, for which the matrix $\Vh$ has to satisfy the condition
\beq
	\Vh^\dg\Vh=\um.
\lbl{eq:V}
\eeq

If the numbers $\Nc_+=\Nc_-=\Nc/2$ of right- and left-moving waves are equal,
in which case the total number $\Nc$ has to necessarily be even, then the square matrix
\beq
	\Vh=\Uh\in\Ux(\Nc/2)
\lbl{eq:Veq}
\eeq
is unitary.
The number of BCs \eqn{VPsi} then also equals $\La=\Nc/2$.
This case of equal numbers $\Nc_\pm$ of right- and left-movers turns out to be the only case of $\Nc_\pm$
when a boundary can be introduced.
We continue with this case in \secr{bc}, while in the remainder of this section
we explore the remaining cases and argue that in the case of unequal numbers $\Nc_\pm$ the boundary cannot be introduced at al.

For unequal numbers of right- and left-moving waves, $\Nc_+>\Nc_-$, {\em one subset} of the solutions to \eq{V}
is the matrix of the form
\beq
	\Vh=\lt(\ba{c} \Uh \\ \nm \ea\rt),
\lbl{eq:Vneq}
\eeq
where $\Uh\in\Ux(\Nc_-)$ is an arbitrary unitary $\Nc_-\tm\Nc_-$ matrix and $\nm$ is an $(\Nc_+-\Nc_-)\tm\Nc_-$ null matrix.
In the BCs \eqn{VPsi}, this corresponds to the vector
\beq
	\Psirh_+[\psih(0)]=
	\lt(\ba{c} \Psirh_+'[\psih(0)] \\ \Psirh_+''[\psih(0)] \ea\rt)
\lbl{eq:Psirhsplit}
\eeq
being split into two parts, with the first $\Nc_-$ components related as
\beq
	\Psirh'_+[\psih(0)]=\Uh\Psih_-[\psih(0)],
\lbl{eq:bc>'}
\eeq
and the last $\Nc_+-\Nc_-$ components nullified:
\beq
	\Psirh''_+[\psih(0)]=\nm,
\lbl{eq:bc>''}
\eeq
where $\nm$ is the null vector of size $\Nc_+-\Nc_-$.

This splitting of the space of $\Psirh_+[\psih(0)]$
into subspaces of $\Nc_-$ and $\Nc_+-\Nc_-$ dimensions can be arbitrary and, by its very definition,
can be parameterized by a Grassmannian $\x{Gr}(\Nc_-,\Nc_++\Nc_-)$.
All possible solutions to \eq{V} are covered by the following family
\beq
	\Vh=\Th^\dg\lt(\ba{c} \Uh \\ \nm \ea\rt)\Th
\lbl{eq:Vneqall}
\eeq
with an arbitrary unitary matrix $\Th$ in which, however, in accord with the notion of Grassmannian,
rotations within the chosen subspaces are redundant, since they all are already spanned by all possible $\Uh$
in the subspace of $\Psirh_+'[\psih(0)]$ and the vector $\Psirh_+''[\psih(0)]$ of the other subspace is nullified.

\subsection{Admissible boundary conditions and Hilbert spaces}

For any numbers $\Nc_\pm$ of right- and left-movers, we have determined above the families of all BCs
with the minimal sufficient number $\La=\max(\Nc_+,\Nc_-)$
that specify the Hilbert spaces over which the current at the boundary is nullified identically.
It could therefore seem that for any $\Nc_\pm$ these BCs describe a quantum-mechanical system with a well-defined boundary.
This is, however, not true.

As we demonstrate in \secr{symmvsherm}, the distinction between the cases of $\Nc_\pm$ for which the boundary is well-defined or not
corresponds precisely to the distinction between self-adjoint (hermitian) and only symmetric (but not self-adjoint) Hamiltonian operators.
In term of the latter, this point has previously been understood.
In Ref.~\ocite{ReedSimon}, it is formulated as ``it is only self-adjoint operators that may be exponentiated'',
i.e., the exponential operator $\ex^{-\ix\Hh t}$, describing the time evolution of the Schr\"odinger equation \eqn{scheq}, is well-defined.
The proof of this statement is rather mathematical.
Although this argument is definitely sufficient (upon establishing the correspondence in \secr{symmvsherm}),
here we also present alternative (but apparently related)
arguments and explanations that are perhaps less mathematically rigorous but more practical and physical.

These arguments are based on the properties of the spectrum:
if the Hamiltonian operator is only symmetric but not self-adjoint (hermitian), its set of eigenvectors is {\em incomplete}.
A system of functions is called complete within a Hilbert space if any function from it can be expanded as a linear combination
of these functions (with appropriate mathematical requirements of convergence of the series),
i.e., they form a proper basis that spans the whole Hilbert space.

\subsubsection{Consideration of two independent boundaries \lbl{sec:twoboundaries}}

First, we consider a system occupying a finite-size segment $x\in[0,X]$ with two boundaries.
Repeating the same procedure \eqn{dt=0} for such a system,
the condition for the wave-function norm to be conserved is that the total net current through the two boundaries vanishes:
\[
	j[\psih(0)]-j[\psih(X)]=0.
\]
There are two subcases here: (i) when the currents through the two boundaries vanish individually,
\beq
	j[\psih(0)]=0,\spc j[\psih(X)]=0,
\lbl{eq:jind=0}
\eeq
and (ii) when the individual currents are nonzero and therefore have to be equal,
\beq
	j[\psih(0)]=j[\psih(X)]\neq 0.
\eeq

The case (ii) is relevant to physical systems in which the coordinates $x=0$ and $x=X$
actually correspond to the same or very close points in the physical real space, like a closed loop with a junction.
Another often-used special case of (ii) are artificial periodic BCs,
where the wave function (for a quadratic-in-momentum Hamiltonian) at $x=0$ and $x=X$ is taken equal (possibly up to a phase factor);
these facilitate calculations of extensive quantities (proportional to the volume of the system) in statistical physics,
for which boundary effects are negligible.

When the points $x=0$ and $x=X$ are truly well-separated in the physical real space
(e.g., when the coordinate $x$ describes the actual quasi-1D geometry),
there is no physical reason for the case (ii) and the currents should vanish individually [\eq{jind=0}].
We refer to this case (i) as the case of {\em independent boundaries}.

For such a natural finite-size system with independent boundaries,
described by the found BCs [\eqs{Vneq}{Vneqall}] at each boundary with generally different matrices $\Vh_0$ and $\Vh_X$,
the problem with these BCs (even though they do nullify the individual currents) in the case of unequal numbers $\Nc_+>\Nc_-$
of right- and left-movers becomes apparent by simple counting of the degrees of freedom.
Consider a stationary Schr\"odinger equation
\beq
	\Hh(\ph)\psih(x)=\e\psih(x)
\lbl{eq:scheqstat}
\eeq
for such a system. The general solution to it is a linear combination of $\Nc=\Nc_++\Nc_-$ linearly independent particular solutions.

For $\Nc_+=\Nc_-=\Nc/2$, the $\La=\Nc/2$ BCs at each of the two boundaries $x=0$ and $x=X$
will produce a system of $\Nc$ linear homogeneous relation for $\Nc$ coefficients of the general solution.
For an arbitrary energy $\e$, this system is nondegenerate and there no nontrivial solutions.
At some energies $\e$, the determinant of the system turns to zero, and there are nontrivial solutions,
which are the eigenstates of the discrete spectrum of the finite-size system.
This is a standard situation, in which one can expect this eigenvector set to be complete, i.e. to span the whole Hilbert space specified by the BCs.

On the other hand, for $\Nc_+>\Nc_-$, the $\Nc_+$ BCs \eqn{Vneq}
at each boundary will produce $2\Nc_+$ linear homogeneous relations for $\Nc=\Nc_++\Nc_-$ coefficients of the general solution.
The excess of $2\Nc_+-(\Nc_++\Nc_-)=\Nc_+-\Nc_->0$ of relations will lead to the absence of the eigenstate solutions:
of course, there cannot be more that $\Nc$ {\em linearly independent} relations between $\Nc$ coefficients,
but this excess will ensure that there will be no accidental degeneracies of the system as a function of $\e$
since the number of linearly independent relations will never drop below $\Nc$.
As a result, for the system with two independent boundaries and $\Nc\neq\Nc_-$, there is no spectrum at all, the eigenvector set is empty.

We arrive at the general conclusion that, for a finite-size system with two independent boundaries,
admissible BCs nullifying the probability currents at each boundary
exist only in the case of equal numbers $\Nc_+=\Nc_-=\Nc/2$ of right- and left-moving waves (in which case $\Nc$ is necessarily even).

Moreover, we realize that there must be ``just the right'' number of BCs.
For equal numbers $\Nc_+=\Nc_-$, this number is the minimal sufficient number $\La=\Nc_+=\Nc_-$ to nullify the current:
less BCs do not nullify the current; more BCs $\La>\Nc_+=\Nc_-$ will keep the current nullified,
but will also nullify the wave-function solution identically for a system with two independent boundaries, via the same argument as above.

For unequal numbers $\Nc_+\neq\Nc_-$, there is no ``right'' number of BCs:
the minimal sufficient number $\La=\max(\Nc_+,\Nc_-)$ of BCs that nullify the current
is already too many, as it leads to the vanishing of the whole wave-function solution for a system with two independent boundaries.
Considering even more (linearly independent) BCs $\La>\max(\Nc_+,\Nc_-)$ (restricting the Hilbert space further)
will only keep the wave-function solution nullified.
This impossibility to introduce a boundary for unequal $\Nc_\pm$ is also consistent with the physical interpretation
of the BCs we provide in \secr{interpretation}.

\subsubsection{Half-infinite system with a linear-in-momentum decoupled Hamiltonian \lbl{sec:halfinfinite}}

Next, for a half-infinite system $x\geq0$, we explicitly prove that the eigenvector set is incomplete for $\Nc_+>\Nc_-$
for the following linear-in-momentum Hamiltonian
\beq
	\Hh(\ph)=\hh_1\ph=\lt(\ba{cc}  \hh_{1+} & \nm \\ \nm& \hh_{1-} \ea\rt)\ph.
\lbl{eq:H10}
\eeq
The wave function
\[
	\psih(x)
	=\lt(\ba{c} \psih_+(x) \\ \psih_-(x) \ea\rt)
	=\lt(\ba{c} \psih_+'(x) \\ \psih_+''(x) \\ \psih_-(x) \ea\rt)
\]
consists of the parts $\psih_\pm(x)$ of sizes $\Nc_\pm$, respectively,
and $\psih_+(x)$ is further split into the subparts $\psih_+'(x)$ and $\psih_+''(x)$ of sizes $\Nc_-$ and $\Nc_+-\Nc_-$, respectively.
The energy matrix $\hh_0$ is assumed absent.
The velocity matrices $\hh_{1+}$ and $\hh_{1-}$ are assumed diagonal (such basis always exists, see \secr{interpretation})
with positive $h_{1+,\nu_+}>0$ and negative $h_{1-,\nu_-}<0$ eigenvalues, respectively.
In this case, the chiral right- and left-moving modes
$\Psir_{\pm,\nu_\pm}[\psih(x)]\equiv\psirh_{\pm,\nu_\pm}(x)=\sq{|h_{1\pm,\nu_\pm}|}\psi_{\pm,\nu_\pm}(x)$
are the wave-function components themselves
and the BCs \eqsn{bc>'}{bc>''} read
\beq
	\psirh_+'(0)=\Uh\psirh_-(0),
\lbl{eq:bc>'1}
\eeq
\beq
	\psirh_+''(0)=\nm.
\lbl{eq:bc>''1}
\eeq
Looking at the eigenvalue problem \eqn{scheqstat},
since all basis states are decoupled in the Hamiltonian \eqn{H10},
the ``remainder'' part, satisfying the BCs \eqn{bc>''1},
has to be nullified identically in the whole system for any eigenstate: $\psih_+''(x)\equiv\nm$, $x\geq0$.
Therefore, the same has to hold for the general wave function belonging to the space spanned by the eigenvectors,
\[
	\psih(x)=\lt(\ba{c} \psih_+'(x) \\ \nm \\ \psih_-(x) \ea\rt), \spc x\geq0,
\]
which means that the set of eigenvectors satisfying the BCs that do nullify the current is incomplete.

This result is intuitively clear: uncompensated chiral modes can be nullified at the boundary only if they are nullified in the whole system.
This removes this part of the wave function from all eigenstates, which makes their set incomplete.
We expect a similar effect for the Hamiltonian of the most general form.
This is the part that we leave without an explicit proof here,
since all other presented arguments suggest that this is indeed the case.

We arrive at the same conclusion for the half-infinite system as for the system with two independent boundaries:
the eigenvector set within the Hilbert space specified by the BCs that nullify the current at the boundary
is complete only for equal numbers $\Nc_+=\Nc_-$ of right- and left-moving mode and incomplete for unequal numbers $\Nc_+\neq\Nc_-$.
Clearly, if the eigenvector set of a Hamiltonian is not complete in a given Hilbert space,
such model cannot describe a well-defined quantum-mechanical system
[Note that the problem here is with the boundary (the Hilbert space), not the bulk (Hamiltonian):
an infinite system without boundaries described by the Hamiltonian with any $\Nc_\pm$ {\em is} well-defined.]
In particular, in a well-defined quantum-mechanical system, it is implied that
any wave function $\psih_0(x)$ from the Hilbert space in question can be used as an initial wave function
in the time-evolution (Cauchy) problem: $\psih(x,t_0)=\psih_0(x)$.
Since the eigenvector set is incomplete, neither the initial wave function $\psih_0(x)$,
nor the time-evolved wave function $\psih(x,t)$ can be expanded in terms of the eigenvector set,
which spans only a smaller subspace of the Hilbert space in question. Such a model is clearly ill-defined.

\subsubsection{Symmetric versus hermitian Hamiltonian operator \lbl{sec:symmvsherm}}

It turns out that the above-demonstrated distinction between which BCs that do nullify the current are admissible and which are not
based on the completeness of the eigenvector set coincides with the
distinction between a self-adjoint (hermitian) and a only symmetric (but not hermitian) Hamiltonian operator.

The constraint \eqn{dt=0} defines a {\em symmetric} Hamiltonian operator $\Hh(\ph)$ over some Hilbert space $\Upsilon$:
i.e., if the Hamiltonian $\Hh(\ph)$ and the Hilbert space $\Upsilon$ are such that \eq{dt=0} is satisfied identically for any $\psih(x)\in\Upsilon$,
then $\Hh(\ph)$ is said to be symmetric over that Hilbert space $\Upsilon$.
We stress again that this is a constraint on {\em both} the form of the Hamiltonian $\Hh(\ph)$ and the Hilbert space $\Upsilon$.
Therefore, by definition, the procedure of finding the BCs that nullify the current carried out above
is simultaneously the procedure of finding the Hilbert spaces $\Upsilon$ over which the Hamiltonian $\Hh(\ph)$ is symmetric.

A {\em self-adjoint} operator is a special case of a symmetric operator satisfying additional, more subtle requirements.
In physical literature, the distinction between a self-adjoint and symmetric Hamiltonian operator
is often overlooked and the operator is called {\em hermitian} regardless, see the last paragraph of this section.
In some mathematical literature~\cite{ReedSimon},
a the term {\em hermitian} is made synonymous to {\em symmetric}, but not to {\em self-adjoint}.
Here, we will adopt a terminology where {\em hermitian} will be synonymous to {\em self-adjoint}.
This seems more meaningful, since self-adjoint operators are the ones that describe a well-defined quantum-mechanical system,
while those that are only symmetric (but not self-adjoint) do not.

We provide a more practical definition of a hermitian operator, which may be less rigorous than in mathematical literature,
but which can immediately be used to check if the operator is hermitian.
To define a hermitian operator, two different wave functions $\psih_{a,b}(x)$ have to necessarily be considered.
Suppose $\Upsilon_b$ is some Hilbert space for $\psih_b(x)$ (specified by some set of BCs),
over which the current $j[\psih_b(0)]$ is nullified, and hence the Hamiltonian is symmetric over $\Upsilon_b$.
One looks for the {\em largest} Hilbert space $\Upsilon_a$ for $\psih_a(x)$, (i.e., specified by the minimal number of BCs) over which the condition
\beq
	\lan\psih_a,\Hh\psih_b\ran=\lan\Hh\psih_a,\psih_b\ran
\lbl{eq:Hhermdef}
\eeq
is satisfied identically, i.e., for all $\psih_a(x)\in\Upsilon_a$ and $\psih_b(x)\in\Upsilon_b$.
Since the Hamiltonian is symmetric over $\Upsilon_b$, the space $\Upsilon_a $ at least includes $\Upsilon_b$.
If these spaces actually happen to be same, $\Upsilon_a=\Upsilon_b$, then the Hamiltonian is said to be hermitian over this one space.
Due to \eqsd{dtab}{dtrhoab}, as for a single wave function [\eq{d=qj}], the relation
\beq
	\lan\psih_a,\Hh\psih_b\ran-\lan\Hh\psih_a,\psih_b\ran
	=\ix j[\psih_a(0),\psih_b(0)]
\lbl{eq:dab=j}
\eeq
holds for the Hamiltonian satisfying the standard bulk requirement \eqn{Hherm};
therefore, the condition \eqn{Hhermdef} being satisfied is equivalent
to the current $j[\psih_a(0),\psih_b(0)]=0$ for two wave functions being nullified at the boundary.

Intuitively, hermiticity means a ``balance'' between the Hilbert spaces $\Upsilon_a$ and $\Upsilon_b$.
The smaller the chosen space $\Upsilon_b$ is, the larger $\Upsilon_a$ will be, and vice versa.
By taking larger spaces $\Upsilon_b$, one can work towards the hermitian Hamiltonian.
However, if the chosen space $\Upsilon_b$ is already the largest possible over which the current $j[\psi_b(0)]$ is nullified
(i.e., over which the Hamiltonian is symmetric) and the other space $\Upsilon_a$ still happens to be larger,
then there is no Hilbert space at all over which the Hamiltonian could be hermitian;
the largest possible space $\Upsilon_b$ is still too constrained.

We now use this definition to check the Hilbert spaces specified by the derived BCs \eqn{VPsi}
with \eqs{bc>'}{bc>''} for $\Nc_+>\Nc_-$ and \eq{Veq} for $\Nc_+=\Nc_-$ for the hermiticity of the Hamiltonian;
in the definition above, $\Upsilon_b$ are now such Hilbert spaces.
The formulas of \secr{jdiagonalization} presenting the current in the universal normalized diagonal form
also hold for the sequilinear form of the current for two different wave functions:
\beq
	j[\psih_a(0),\psih_b(0)]
	=\Psirh_+^\dg[\psih_a(0)]\Psirh_+[\psih_b(0)]-\Psirh_-^\dg[\psih_a(0)]\Psirh_-[\psih_b(0)].
\lbl{eq:jabdn}
\eeq

For unequal numbers $\Nc_+>\Nc_-$ of right- and left-movers, using the splitting \eqn{Psirhsplit}, we have
\beq
	j[\psih_a(0),\psih_b(0)]
	=\Psirh_+'^\dg[\psih_a(0)]\Psirh_+'[\psih_b(0)]+\Psirh_+''^\dg[\psih_a(0)]\Psirh_+''[\psih_b(0)]-\Psirh_-^\dg[\psih_a(0)]\Psirh_-[\psih_b(0)].
\lbl{eq:jab}
\eeq
For $\psih_b(x)\in\Upsilon_b$ satisfying the BCs \eqsn{bc>'}{bc>''}, of which there are $\Nc_+$ relations,
no restriction on $\Psirh''_+[\psih_a(0)]$ is required to nullify the whole current.
The {\em largest} space $\Upsilon_a$ over which the current \eqn{jab} is nullified is therefore
specified by the minimal number $\Nc_-$ of BCs of the form
\[
	\Psirh'_+[\psih_a(0)]=\Uh\Psih_-[\psih_a(0)],
\]
while $\Psirh''_+[\psih_a(0)]$ can be arbitrary.
Hence, this Hilbert space $\Upsilon_a$ of $\psih_a(x)$ is larger than $\Upsilon_b$ of $\psih_b(x)$ ($\Nc_-$ versus $\Nc_+$ BCs, respectively)
and the Hamiltonian is not hermitian over $\Upsilon_b$, only symmetric.
This, of course, also applies to the general form \eqn{Vneqall}.
Since $\Upsilon_b$ is the largest subspace (specified by the minimal sufficient number $\La=\Nc_+$ of BCs)
over which the current $j[\psih_b(0)]$ is nullified,
there is no Hilbert space at all over which the Hamiltonian would be hermitian for $\Nc_+\neq\Nc_-$.

For equal numbers $\Nc_+=\Nc_-$ of right- and left-movers, the remainder part $\Psirh''[\psih_a(0)]$ is absent,
and the largest space $\Upsilon_a$ of $\psih_a(x)$ coincides with the space $\Upsilon_b$ of $\psih_b(x)$, specified by the BCs \eqn{bc}.
Hence, the Hamiltonian over such Hilbert space is not only symmetric but also hermitian.

Therefore, we recognize that the cases of BCs nullifying the current that are admissible (only $\La=\Nc_+=\Nc_-$)
or not (any $\La>\Nc_+=\Nc_-$ for $\Nc_+=\Nc_-$ and any $\La\geq\max(\Nc_+,\Nc_-)$ for $\Nc_+\neq\Nc_-$)
based on the completeness of the eigenvector set (\secsr{twoboundaries}{halfinfinite})
precisely coincide with the cases of Hilbert spaces over which the symmetric Hamiltonian is also hermitian or not.

We conclude that the question of which BCs (and the Hilbert spaces they specify) are admissible
in the sense that they represent a system with a well-defined boundary
turns out to be more subtle than the mere nullification of the current at the boundary.
In Ref.~\ocite{Ahari}, this part of analysis has not been presented.

In physical literature, this subtle but crucial distinction between symmetric and hermitian (self-adjoint) operators
is oftentimes overlooked and the definition \eqn{Hsymmdef} of symmetric operator
is used as the (inaccurate) definition of a hermitian operator.
This inaccuracy does not lead to erroneous conclusions for systems with $\Nc_+=\Nc_-$, which is typical,
and when the right number $\La=\Nc_+=\Nc_-$ of linearly independent BCs is correctly ``guessed'',
since this is precisely the only case when symmetric and hermitian operators do become equivalent.

\subsubsection{Comment on the relation of the ``self-adjoint extension'' to the problem of general boundary conditions \lbl{sec:selfadjext}}

In some studies~\cite{ReedSimon,Ahari,Seradjeh} of the general BCs for CMs, the notion of {\em ``self-adjoint extension''} is brought up.
The unabbreviated name for the ``self-adjoint extension'' procedure reads ``extension of a symmetric operator to a self-adjoint operator''.
In accord with the definition of a self-adjoint (hermitian) operator given above,
this procedure is realized when one starts with a Hilbert space $\Upsilon_b$ over which the Hamiltonian operator is symmetric,
but it is not the largest such space (there is room ``to extend'') and one tries to find such space; hence the term ``extension''.
Note that the form of the operator is not changed at all during this procedure, only the Hilbert space is; what is ``extended'' is the Hilbert space.
In terms of BCs, in the self-adjoint-extension procedure, one therefore starts with possibly ``too many'' BCs that do nullify the current
and works towards the number of BCs that is less than the initial one.

On the other hand, in the natural formulation of the BCs problem for a quantum-mechanical system, as presented in \secr{herm},
one starts with the embedding Hilbert space $\Upsilon_0$ of wave functions with no BCs at all, and recognizes that such restrictions are necessary.
One then finds the BCs with the {\em minimal sufficient} number to nullify the current.
Therefore, since $\Upsilon_b$ is already by construction the largest space over which the current is nullified and the Hamiltonian is symmetric,
one can only check if the Hamiltonian is also hermitian over this Hilbert space, as done in \secr{symmvsherm};
it is already known that there is ``nowhere to extend''.

Therefore, while the end goal of the two procedures is the same, these are opposite starting points
and the starting point of the ``self-adjoined extension'' procedure is
rather artificial for the problem at hand: some BCs that do nullify the current, but there are too many of them.
Therefore, we believe that specifically the use of the term ``extension''
is not relevant to the natural formulation of the problem of general BCs and is somewhat misleading;
``finding admissible Hilbert spaces over which the Hamiltonian becomes hermitian''
seem like an accurate and more appropriate description of the procedure.

\subsection{Standardized universal form of the general boundary conditions \lbl{sec:bc}}

Summarizing the results of the previous subsections,
for a system described by a continuum model with the translation-symmetric Hamiltonian $\Hh(\ph)$ of the most general form \eqn{H},
a boundary can be introduced {\em only for equal numbers}
\[
	\Nc_+=\Nc_-=\f{\Nc}2
\]
of positive and negative eigenvalues of the current matrix (right- and left-moving waves, respectively, in the geometry of \figr{system}).
Only in this case admissible Hilbert spaces, over which the Hamiltonian is not only symmetric
[which is equivalent to the current at the boundary being nullified, \eqs{Hsymmdef}{d=qj}]
but also hermitian [\eq{Hhermdef}], exist.
The family of general BCs specifying all such admissible Hilbert spaces reads
\beq
	\Psirh_+[\psih(x=0)]=\Uh\Psirh_-[\psih(x=0)].
\lbl{eq:bc}
\eeq
This is a matrix form combining $\La=\Nc/2$ linearly independent homogeneous scalar relations
between the wave-function components and their derivatives.
All possible BCs are parameterized by all possible unitary matrices
\beq
	\Uh\in\Ux(\Nc/2)
\lbl{eq:U}
\eeq
of order $\Nc/2$, with $(\Nc/2)^2$ real parameters.
This reproduces the central result of Ref.~\ocite{Ahari}
(although no substantiation for it was provided therein, most importantly, why this is the only well-defined case).

The two most valuable properties of this form of general BCs are:

(i) The universal standardized form of the BCs \eqn{bc}.
The current [\eqs{jd}{jdn}] and, as a consequence, the BCs \eqn{bc}
are presented in the universal forms in terms of the linear combinations $\Psir_{\pm,\nu_\pm}[\psih(0)]$ [\eq{Psir}]
of the wave-function components and their derivatives $\ph^n\psi_m(0)$ [\eq{pnpsi}]
that diagonalize and normalize the current.
These forms are fully characterized for {\em any} Hamiltonian
just by the numbers $\Nc_\pm$ of positive and negative eigenvalues of the current matrix \eqn{Jdef},
interpreted as the numbers or right- and left-moving waves.
The BCs can then be simply expressed in terms of the original degrees of freedom \eqn{pnpsi} via diagonalization formulas \eqsn{Psipm}{Psir}.

(ii) Nonredundant one-to-one (bijective) parametrization of the general BCs:
one instance of $\Uh$ provides one distinct set of possible BCs, specifying one admissible Hilbert space;
possible redundancies of BCs due equivalence transformations or additional linearly dependent relations (see \appr{nonredundance}),
which do not change the Hilbert space, are eliminated in this formalism by the imposed structure.

The physical meaning behind the general BCs will be discussed in \secr{realization}.

\section{Physical interpretation of the general boundary conditions as a scattering process \lbl{sec:interpretation}}

\begin{figure}
\includegraphics[width=.23\textwidth]{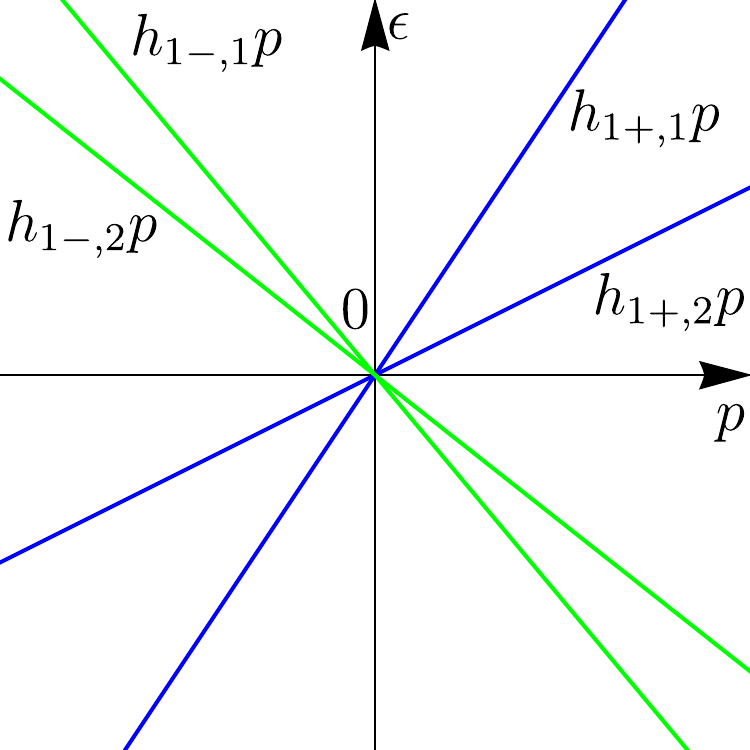}
\caption{
The bulk spectrum of the part $\hh_1\ph$ of the linear-in-momentum ($N=1$) Hamiltonian $\Hh_1(\ph)=\hh_1\ph+\hh_0$
for a multi-component wave function, illustrated with the case of $\Nc_\pm=2$ right- and left-moving waves (modes).
The ``energy'' matrix $\hh_0$ is not involved in the current [\eq{J1gen}] and hence, in the general BCs \eqn{bc}.
The momentum-independent eigenstates of the linear part $\hh_1 p$ can be directly identified as right- and left-moving waves
and the interpretation of the general BCs \eqn{bc} as a scattering process between them [\secr{interpretation}, \figr{system}(b)]
for the linear-in-momentum Hamiltonian becomes particularly transparent.
}
\lbl{fig:H1}
\end{figure}

The form \eqn{bc} of the general BCs immediately suggests a natural physical interpretation, which we now provide.
Namely, as already mentioned in \secr{jdiagonalization},
the projections $\Psi_{\pm,\nu_\pm}[\psih(x)]$ [\eq{Psipm}]
of the vector $\Psih[\psih(x)]$ onto the eigenvectors $\sh_{\pm,\nu_\pm}$ of the current matrix $\Jh$ [\eq{Jdef}]
with positive and negative eigenvalues $\Jc_{\pm,\nu_\pm}\gtrless0$ can be interpreted as chiral ``waves'' (or ``modes'') propagating
in the right and left directions, respectively, in the geometry of \figr{system},
according to the signs of their contributions to the diagonal form of the current [\eqs{jd}{jdn}].
For the $x\geq 0$ system, these are the waves reflected from and incident upon the boundary $x=0$, respectively.
The general BCs \eqn{bc} can therefore be seen as a scattering process between these waves at the boundary,
with $\Uh$ being the scattering matrix, as illustrated in \figr{system}.

The interpretation is most transparent in the case of the linear-in-momentum ($N=1$) Hamiltonian
\beq
	\Hh(\ph)=\hh_0+\hh_1\ph,
\lbl{eq:H1gen}
\eeq
with the ``energy'' $\hh_0$ and ``velocity'' $\hh_1$ matrices, \figr{H1}. In this case [see \eqs{Psi}{J} in the next section], the vector
\[
	\Psih[\psih(x)]=\psih(x)
\]
in \eq{Jdef} is equal to the wave-function vector \eqn{psi} itself and the current matrix
\beq
	\Jh=\hh_1
\lbl{eq:J1gen}
\eeq
is given by the velocity matrix.
The number $\Nc=NM=M$ of the degrees of freedom equals to the number of the wave-function components.
The eigenvector $\sh_{\pm,\nu_\pm}$ and eigenvalues $\Jc_{\pm,\nu_\pm}=h_{1\pm,\nu_\pm}$
of the current matrix \eqn{J1gen} are therefore those of the velocity matrix $\hh_1$.
The boundary is well-defined only for equal $\Nc_+=\Nc_-=M/2$
and the general BCs have the form \eqn{bc} with the projections
\[
	\Psi_{\pm,\nu_\pm}[\psih(x)]=\sh^\dg_{\pm,\nu_\pm}\psih(x),
\]
\[
	\Psir_{\pm,\nu_\pm}[\psih(x)]=\sq{|h_{1\pm,\nu_\pm}|}\Psi_{\pm,\nu_\pm}[\psih(x)].
\]
As already mentioned in \secr{j}, since the energy matrix $\hh_0$ does not enter the current \eqn{J1gen},
the BCs \eqn{bc} are the same for any $\hh_0$, including the absent one.
For $\hh_0=\nm$, the system is gapless; the bulk spectrum consists of the linearly dispersing bands $\eps_{\pm,\nu_\pm}(p)=h_{1\pm,\nu_\pm} p$,
all crossing the point $(p,\e)=(0,0)$, and the eigenvectors $\sh_{\pm,\nu_\pm}$ of the current matrix \eqn{J1gen}
coincide with the respective {\em momentum-independent} eigenvectors of the Hamiltonian $\Hh(p)=\hh_1 p$.

For the linear-in-momentum Hamiltonian \eqn{H1gen},
the right- and left-moving waves can therefore be identified as the bulk eigenstates of its linear part $\hh_1\ph$.
The interpretation of the general BCs \eqn{bc} as a scattering process is very natural in this case.
For momentum order $N>1$ higher than linear, this interpretation is a little less intuitive,
but nonetheless still meaningful according to the structure of the current [\eqs{jd}{jdn}] and the resulting BCs \eqn{bc}.

From this scattering-process interpretation, one can also better understand the conditions
for when the boundary can be introduced and admissible BCs exist:
from the causality argument, the set of reflected waves should be completely determined by the set of incident waves
(and the reflected and incident waves can also be interchanged by time reversal).
The derived requirements that the numbers $\Nc_\pm$ of the reflected and incident waves must be equal,
$\Nc_-=\Nc_+=\Nc/2$, and the number $\La=\Nc/2$ of BCs (relations between them) must also be equal to that number
are therefore fully consistent with the causality argument.
If the numbers $\Nc_\pm$ of right- and left-moving waves are unequal, there are ``uncompensated'' waves;
physically, this means that such a system cannot be terminated in 1D.
The prime example would be an edge of a quantum Hall system, having, in the simplest case,
one edge-state mode, described by the Hamiltonian $\Hh(\ph)=h_1 \ph$ with the linearized spectrum at low energies, with $\Nc_+=1$ and $\Nc_-=0$.
Physically, as is well-known from the quantum Hall physics, attempting to terminate the 1D edge of a 2D sample by ``cutting'' the sample
will force the edge state to propagate along the newly created edge.
This is also the simplest case illustrating the general considerations of \secr{symmvsherm}.
The largest and only Hilbert space $\Upsilon_b$, over which the Hamiltonian is symmetric, i.e., the current $j[\psi_b(0)]=h_1\psi_b^*(0)\psi_b(0)$
is nullified identically, is that of the wave functions satisfying the BC $\psi_b(0)=0$.
The current $j[\psi_a(0),\psi_b(0)]=h_1\psi_a^*(0)\psi_b(0)\equiv 0$ is then identically zero
without any restrictions on $\psi_a(0)$. Therefore, there is no Hilbert space over which the Hamiltonian could be hermitian.

\section{Structure of the current matrix and justification of the formalism for top momentum order $N>1$ \lbl{sec:justification}}

In the derivation of the general BCs presented in Ref.~\ocite{Ahari} and reproduced with more details in \secr{jdiagonalization},
the diagonalization of the current matrix $\Jh$ [\eq{Jdef}] was assumed to be a well-defined procedure.
However, as we now show, upon closer inspection, there is an important subtlety
in this procedure for the top momentum order $N>1$ higher than linear
that needs to be addressed. Simultaneously, we explore the structure of the current matrix.

We notice that for any Hamiltonian with $N>1$ presenting the current $j[\psih(x)]$ in the matrix form \eqn{Jdef}
and subsequent diagonalization of the current matrix $\Jh$ inevitably involves introducing an auxiliary fictitious length scale $l$.
Indeed, these operations are well-defined only if the components of the vector $\Psih[\psih(x)]$
have the same physical dimension and, as a result, the elements of $\Jh$ have the same physical dimension.
Only commensurable quantities of the same physical dimension can sensibly be treated as components of one vector or matrix,
upon which linear transformations may be performed. Consider a positive length parameter
\beq
	l>0.
\lbl{eq:l}
\eeq
Then all derivatives
\[
	\prh^n\psi_m(x)=(-\ix l\pd_x)^n\psi_m(x)
\]
of different orders $n$ with the dimensionless momentum operator
\beq
	\prh=l\ph
\lbl{eq:pr}
\eeq
have the same physical dimension and may be joined into one vector $\Psih[\psih(x)]$.

\subsection{Case of equal top-momentum orders $N_m=N$}

We first consider the technically simplest case of all wave-function components $\psi_m(x)$, $m=1,\ldots,M$,
having the same $N_m=N$ top momentum order in the Hamiltonian $\Hh(\ph)$ [\eq{H}], in which case $\Nc=NM$ [\eq{Nc}].
We introduce the vector as
\beq
	\Psih[\psih(x)]=\lt(\ba{c} \psih(x) \\\prh\psih(x)\\ \prh^2\psih(x) \\\ldots\\ \prh^{N-1}\psih(x)\ea\rt).
\lbl{eq:Psi}
\eeq
The current is then presented in the form \eqn{Jdef} with the hermitian current matrix of the block-triangular structure
\beq
	\Jh=\lt(\ba{ccccc}
		\Jh_1&\Jh_2&\Jh_3&\ldots&\Jh_N\\
		\Jh_2&\Jh_3&\ldots&\Jh_N&\nm\\
		\Jh_3&\ldots&\Jh_N&\nm&\ldots\\
		\ldots&\Jh_N&\nm&\ldots&\nm\\
		\Jh_N&\nm&\ldots&\nm&\nm\\
		\ea\rt)
	,\spc \Jh^\dg=\Jh,
\lbl{eq:J}
\eeq
where
\beq
	\Jh_n=\f{\hh_n}{l^{n-1}}, \spc \Jh_n^\dg=\Jh_n, \spc n=1,\ldots,N,
\lbl{eq:Jn}
\eeq
and $\nm$ is the $M\tm M$ zero matrix.
Due to the length scale $l$ in $\Psih[\psih(x)]$ [\eq{Psi}], the matrices $\Jh_n$,
describing the contributions to the current from each momentum order $\hh_n \ph^n$ in the Hamiltonian $\Hh(\ph)$,
contain the factors $1/l^{n-1}$, which ensures that they all have the same physical dimension of velocity.
This way, the properly introduced current matrix $\Jh=\Jh(\hh_1,\ldots,\hh_N;l)$ and its eigenvalues
necessarily depend on the fictitious length scale $l$, which may be changed at will.

We point out that the zero-order part $\hh_0$ of the Hamiltonian,
the energy matrix representing various energy-level shifts of and couplings between the basis states of the wave function,
never enters the current \eqn{j} and hence does not affect at all the form \eqn{bc} of the general BCs.
The BCs are the same for any $\hh_0$, in particular, for absent $\hh_0=\nm$.

The current matrix $\Jh$ [\eq{J}] for equal top momentum orders $N_m=N$ has a special block-triangular structure:
there are no terms on one side of the anti-diagonal and there is the same $\Jh_n$ matrix on each sub-anti-diagonal on the other side.
This is the consequence of the symmetrized form \eqn{jn} of the current.
This structure leads to the key property of the eigenvalues of the current matrix $\Jh$, which is crucial for the BCs formalism to be well-defined.
This property is based on the following theorem.

{\em Theorem: For equal top-momentum orders $N_m=N$, the current matrix $\Jh$ [\eq{J}] is degenerate if and only if the top-momentum-order matrix $\hh_N$
in the Hamiltonian $\Hh(\ph)$ [\eq{H}] is degenerate.}

\subsubsection{Proof of the theorem \lbl{sec:Jtheorem}}

Proof of {\em If $\Jh$ is degenerate, then $\hh_N$ is degenerate.}

Consider the matrix equation
\beq
	\Jh\Psih=\nm
\lbl{eq:Jheq}
\eeq
for the vector
\[
	\Psih=\lt(\ba{c} \Psih_0\\\Psih_1\\\ldots\\\Psih_{N-1}\ea\rt),
\]
consisting of the subvectors $\Psih_n$ ($n=0,\ldots,N-1$) of size $M$.
In this section, $\nm$ are null vectors of the respective sizes.
Using the structure \eqn{J} of the current matrix, \eq{Jheq} is equivalent to $N$ matrix equations
\[
	\Jh_1\Psih_0+\ldots+\Jh_N\Psih_{N-1}=\nm,
\]
\[
	\ldots
\]
\beq
	\Jh_{N-1}\Psih_0+\Jh_N\Psih_1=\nm,
\lbl{eq:JPsi1}
\eeq
\beq
	\Jh_N\Psih_0=\nm.
\lbl{eq:JPsi0}
\eeq

Suppose $\Jh_N=\hh_N/l^{N-1}$ is nondegenerate. It then follows from \eq{JPsi0} that $\Psih_0=\nm$.
\eq{JPsi1} then reduces to $\Jh_N\Psih_1=\nm$, from which it follows that $\Psih_1=\nm$.
Repeating this procedure, we obtain that all $\Psih_n=\nm$ ($n=0,\ldots,N-1$)
vanish, and hence, the matrix equation \eqn{Jheq} has only a trivial solution $\Psih=\nm$.
Therefore, the current matrix $\Jh$ is nondegenerate,
regardless of the values of the lower-momentum-order matrices $\Jh_n=\hh_n/l^{n-1}$, $n=1,\ldots,N-1$.
The statement is proven by contradiction.

Proof of {\em If $\hh_N$ is degenerate, then $\Jh$ is degenerate.}

Suppose $\Psih_{N-1}\neq\nm$ is the eigenvector of $\Jh_N$ with zero eigenvalue, $\Jh_N\Psih_{N-1}=\nm$.
The nonzero vector
\[
	\Psih=\lt(\ba{c} \nm\\...\\\nm\\\Psih_{N-1}\ea\rt)\neq\nm
\]
then gives
\[
	\Jh\Psih=\nm,
\]
and hence, $\Jh$ is degenerate.

\subsection{Case of unequal top-momentum orders $N_m$}

Here, we present a similar key theorem for the case of unequal
top momentum orders $N_m=N$ of the wave-function components $m=1,\ldots,M$.
Not to overburden the presentation, we consider the case of different $N_m$ for $N=3$; generalization to higher $N$ is straightforward.

We first group the components of the wave function \eqn{psi}
\[
	\psih(x)=\lt(\ba{c} \psih_1(x)\\\psih_2(x)\\\psih_3(x) \ea\rt)
\]
into subvectors $\psih_{1,2,3}(x)$ that have the top momentum orders $N_m=1,2,3$, respectively.
We assume that for each $N_m$ there is at least one component, i.e.,
the sizes $M_{1,2,3}>0$ of the vectors $\psih_{1,2,3}(x)$ are nonzero; $M_1+M_2+M_3=M$.
In this block basis, the most general form of the Hamiltonian \eqn{H} reads
\beq
	\Hh(\ph)
	=\hh_0+\hh_1\ph+\hh_2\ph^2+\hh_3\ph^3
	=\hh_0
	+\lt(\ba{ccc}
	\hh_1^{11} & \hh_1^{12} & \hh_1^{13} \\
	\hh_1^{21} & \hh_1^{22} & \hh_1^{23} \\
	\hh_1^{31} & \hh_1^{32} & \hh_1^{33}
	\ea\rt)\ph
	+\lt(\ba{ccc}
	\nm  & \nm & \nm \\
	\nm & \hh_2^{22} & \hh_2^{23} \\
	\nm & \hh_2^{32} & \hh_2^{33}
	\ea\rt)\ph^2
	+\lt(\ba{ccc}
	\nm  & \nm & \nm \\
	\nm & \nm & \nm \\
	\nm & \nm & \hh_3^{33}
	\ea\rt)\ph^3,
\lbl{eq:Huneq}
\eeq
with $(\hh_n^{kk'})^\dg=\hh_n^{k'k}$, $k,k'=1,2,3$,
and $\nm$ being zero matrices of respective orders.

We introduce the vector as
\[
	\Psih[\psih(x)]=\lt(\ba{c} \psih_1(x)\\ \psih_2(x)\\ \psih_3(x)\\ \prh\psih_2(x)\\ \prh\psih_3(x) \\ \prh^2\psih_3(x) \ea\rt).
\]
The current quadratic form $j[\psih(x)]$ [\eq{j}] can then be presented in the matrix form \eqn{Jdef} with this vector and
\beq
	\Jh=
	\lt(\ba{cccccc}
		\Jh_1^{11} & \Jh_1^{12} & \Jh_1^{13} & \nm & \nm & \nm\\
		\Jh_1^{21} & \Jh_1^{22} & \Jh_1^{23} & \Jh_2^{22} & \Jh_2^{23} &\nm\\
		\Jh_1^{31} & \Jh_1^{32} & \Jh_1^{33} & \Jh_2^{32} & \Jh_2^{33} & \Jh_3^{33}\\
		\nm& \Jh_2^{22} & \Jh_2^{23} & \nm & \nm &\nm\\
		\nm& \Jh_2^{32} & \Jh_2^{33} & \nm & \Jh_3^{33} &\nm\\
		\nm& \nm & \Jh_3^{33} & \nm & \nm &\nm\\
	\ea\rt),
	\spc
\lbl{eq:Juneq}
\eeq
\[
	\Jh_n^{kk'}=\f{\hh_n^{kk'}}{l^{n-1}}, \spc k,k'=1,2,3.
\]

In essence, the current matrix \eqn{Juneq} can be obtained from the one \eqn{J}
for equal top momentum orders $N_m=3$ with $\hh_n$ from \eq{Huneq} by discarding the zero rows and columns in the latter,
as well as discarding the respective derivatives of the wave-function components in the vector $\Psih[\psih(x)]$.
This analogous structure of the current matrix \eqn{Juneq}
leads to the key property of its eigenvalues, which follows from the theorem.

{\em Theorem: For unequal top-momentum orders $N_m=1,2,3$ for different wave-function components,
the current matrix $\Jh$ [\eq{Juneq}] is degenerate if and only if
at least one of the top-momentum-order matrices $\hh_1^{11}$, $\hh_2^{22}$, or $\hh_3^{33}$
in the Hamiltonian \eqn{Huneq} is degenerate.}

The proof is analogous to the one for the case of equal $N_m=N$.

\subsection{Sign structure of the current-matrix eigenvalues is independent of $l$}

The degeneracy of the current matrix $\Jh$ as its parameters are spanned
signifies a change of the signs of one or more of its eigenvalues as they cross zero.
Therefore, it follows from the above theorems that the {\em sign structure} of the eigenvalues $\Jc_\nu$, $\nu=1,\ldots,\Nc$,
of the current matrix $\Jh$
is {\em determined solely by the top-momentum-order matrices} in the Hamiltonian: by $\hh_N$ for equal $N_m=N$ and by $\hh_n^{nn}$
[$n=1,2,3$, \eq{Huneq}] for unequal $N_m$.
Namely, the signs of the eigenvalues $\Jc_\nu$ can never be changed by changing any of the lower-order matrices
or, most importantly, by changing the fictitious length-scale parameter $l$ [\eqs{l}{pr}],
even though the eigenvalues $\Jc_\nu$ themselves do depend on them.
This justifies the labelling of the eigenvalues according to their signs as
\[
	\Jc_{+,\nu_+}=\Jc_\nu>0, \spc \nu_+=1,\ldots,\Nc_+,
\]
\[
	\Jc_{-,\nu_-}=\Jc_\nu<0, \spc \nu_-=1,\ldots,\Nc_-,
\]
as well as the labelling of the respective eigenvectors $\sh_{\pm,\nu_\pm}$ and projections $\Psi_{\pm,\nu_\pm}[\psih(x)]$ [\eqs{pnpsi}{Psipm}]
of $\Psih[\psih(x)]$ [\eq{Psi}] onto these eigenvectors.
Consequently, the corresponding characterization of the projections $\Psi_{\pm,\nu_\pm}[\psih(x)]$
as right- and left-moving waves, introduced in \secr{interpretation},
is also determined solely by the top-momentum-order matrices.
An explicit illustration for the fourth-order-in-momentum ($N=4$) Hamiltonian is provided in \appr{JN4} as an example.

This property that the sign structure of the current-matrix eigenvalues is independent of the fictitious length scale $l$
is extremely important as it provides an essential mathematical justification for the procedure of deriving the general BCs \eqn{bc}.
Otherwise, the whole formalism of the general BCs would be ill-defined for Hamiltonians with the top momentum orders $N>1$ higher than linear.

The degeneracy of the top-momentum-order matrix $\hh_N$ (or one of $\hh_n^{nn}$) as its parameters are spanned,
when the current matrix $\Jh$ does become degenerate and the sign changes of its eigenvalues $\Jc_\nu$ do occur,
signifies a qualitative change in the band structure
(which is understandable since the asymptotic band structure at large momenta is determined by them).
Therefore, it is sensible to treat the sectors in the parameter space of $\hh_N$ where it is nondegenerate separately,
as regions where the key qualitative properties of the band structure are preserved.
For example, for the quadratic-in-momentum ($N=2$) Hamiltonian $\Hh(p)=h_2 p^2$ for a one-component ($M=1$) wave function, the current matrix
\beq
	\Jh=\f{h_2}{l}\lt(\ba{cc}0&1\\1&0\ea\rt)
\lbl{eq:J2}
\eeq
with eigenvalues $\pm |h_2|/l$ is nondegenerate in both sectors $h_2>0$ and $h_2<0$;
in each of them, $\Nc_+=\Nc_-=1$ and the boundary and general BC are well-defined, see \secr{N2M1} below.
But at $h_2=0$ the bulk spectrum becomes flat, as the change between the positive and negative curvature occurs.

\subsection{Fictitious length scale $l$ is not an additional parameter of the general boundary conditions\lbl{sec:l}}

Although the fictitious length scale $l$ has to inevitably be introduced into the formalism for top-momentum order $N>1$
and enters the general BCs via the $\Psih[\psih(x);l]$ dependence,
here we explain that it is {\em not} an additional free parameter of the general BCs.

Let us rewrite the general BCs \eqn{bc} with the dependence of the vectors $\Psirh_\pm[\psih(x)]=\Psirh_\pm[\psih(x);l]$ on $l$
explicitly indicated:
\beq
	\Psirh_+[\psih(x=0);l]=\Uh\Psirh_-[\psih(x=0);l].
\lbl{eq:bcl}
\eeq
For a fixed $l$, spanning all possible unitary matrices $\Uh\in\Ux(\Nc/2)$ will span the family of all possible BCs.

If, for a fixed $\Uh$, $l$ is changed to some other value $l'$, one obtains the BCs
\[
	\Psirh_+[\psih(x=0);l']=\Uh\Psirh_-[\psih(x=0);l']
\]
that are not equivalent to \eq{bcl} (they specify a different Hilbert space and cannot be obtained by equivalence transformations).
However, one can always find a unitary matrix $\Uh'=\Uh'(l,l';\Uh)$, which would depend on $\Uh$, $l$, and $l'$,
such that the latter BCs {\em are} equivalent to
\[
	\Psirh_+[\psih(x=0);l]=\Uh'\Psirh_-[\psih(x=0);l]
\]
with the original $l$. This point will also be explicitly illustrated in \secr{N2M1} for the quadratic-in-momentum ($N=2$) one-component ($M=1$) model.

\section{Semi-analytical method of calculating bound states\lbl{sec:method}}

Once the BCs have been derived, the CM of a system with a boundary is fully specified by its Hamiltonian $\Hh(\ph)$ [\eq{H}] and BCs \eqn{bc}
and one can calculate the complete set of eigenstates and their spectrum.
Bound states are the eigenstates that decay into the bulk, with the wave function $\psih(x)\rarr \nm$ as $x\rarr+\iy$ for the $x>0$ half-infinite system.
(The eigenstates of the continuum part of the spectrum for the system with a boundary can, of course, also be calculated;
these can be formulated in terms of a scattering problem.)

Here we present the (semi)-analytical method of calculating the bound states within CMs with BCs.
This method has been used in some previous works, especially for the simplest models;
nonetheless, it seems worthwhile to formulate it clearly and emphasize it,
since other, less efficient methods, like finite-size calculations, are still being commonly used.
An analogous method for lattice models has also been formulated~\cite{Istas2017}.

The method follows directly from the theory of linear differential equations.
Bound states can exist only within the energy gaps of the bulk spectrum
of the respective infinite system, i.e., the spectrum of the matrix $\Hh(p)$ for real $p$.
One first constructs the general solution to the stationary Schr\"odinger equation \eqn{scheqstat}
at a given energy $\e$ within the gaps that decays into the bulk.
Such general solution is a linear combination of particular solutions of the form
\beq
    \chih \ex^{\ix p x},
\lbl{eq:chipart}
\eeq
where the momentum $p$ satisfies the characteristic equation
\beq
	\det[\Hh(p)-\e\um]=0
\lbl{eq:chareq}
\eeq
and $\chih$ is the corresponding nontrivial ``eigenvector'' solution to
\[
    [\Hh(p)-\e\um]\chih=0.
\]

There are $\Nc$ [\eq{Nc}] momentum solutions $p_\al(\e)$ to \eq{chareq}.
For the energy $\e$ within the gap, there are, by construction, no real momentum solutions; each momentum solution has a nonzero imaginary part.
Due to $\Nc_+=\Nc_-=\Nc/2$, there always are $\Nc/2$ momentum solutions with positive and negative imaginary parts.
For the $x>0$ system, only the $\Nc/2$ momentum solutions with a positive imaginary part $\Imx p_\al(\e)>0$ are kept, labeled $\al=1,\ldots,\Nc/2$,
to have decaying particular solutions \eqn{chipart}. The general solution at a given energy $\e$, decaying into the bulk, reads
\beq
	\psih(x;\e)=\sum_{\al=1}^{\Nc/2} c_\al\chih_\al(\e)\ex^{\ix p_\al(\e) x},
\lbl{eq:psigen}
\eeq
where $c_\al$ are arbitrary coefficients.
[It is assumed here that there are indeed $\Nc$ linearly independent eigenvectors $\chih_\al(\e)$ for $\Nc$ momentum solutions $p_\al(\e)$.
In the case of degenerate momentum solutions $p_\al(\e)$, the number of eigenvectors may sometimes be less than the multiplicity.
In that case, enough particular solutions still exists, but their coordinate dependence differs from that of \eq{chipart}.
The adaptation to this case is also straightforward and follows from the theory of differential equations.]

This general solution is then inserted into the general BCs \eqn{bc}. Denoting
\[
	\Psirh_\pm[\chih_\al(\e)\ex^{\ix p_\al(\e) x}]
	=\Psirh_\pm^\al(\e)\ex^{\ix p_\al(\e) x}
\]
according to the action of the momentum operator $\ph$ in $\Psirh_\pm[\psih(x)]$,
we obtain the homogeneous system
\[
	\sum_{\al=1}^{\Nc/2}[\Psirh_+^\al(\e)-\Uh\Psirh_-^\al(\e)]c_\al=\nm
\]
of $\Nc/2$ equations for $\Nc/2$ unknown variables $c_\al$, where $\nm$ is the null vector of size $\Nc/2$.
One can equivalently present this system in the matrix form
\beq
    \Xh(\e)\ch=\nm,
\lbl{eq:eqc}
\eeq
where
\[
	\Xh(\e)=
	\lt(\ba{ccc}
	\Psirh_+^1(\e)-\Uh\Psirh_-^1(\e) &
		\ldots &
		\Psirh_+^{\Nc/2}(\e)-\Uh\Psirh_-^{\Nc/2}(\e)
	\ea\rt)
\]
is an $\Nc/2\tm\Nc/2$ matrix and
\[
    \ch=\lt(\ba{c} c_1\\\ldots\\c_{\Nc/2} \ea\rt)
\]
is the vector of the coefficients.

The problem of finding bound states has been reduced to solving the system \eqn{eqc}
of equations for the coefficients of the general decaying solution \eqn{psigen}.
Generally, for an arbitrary energy $\e$, the matrix $\Xh(\e)$ is nondegenerate and there are no nontrivial solutions for $\ch$.
Nontrivial solutions appear at those energies $\e$ at which the matrix becomes degenerate, i.e.,
\[
	\det\Xh(\e)=0
\]
is the equation for the energy $\e$ of possible bound states.
The solutions to this equation determine the bound-state energies
and the corresponding nontrivial solutions $\ch$ to \eq{eqc} yield their wave functions according \eq{psigen}.

We consolidate the discussion of all the advantages that the formalism of general BCs
and this calculation method provide for the study of bound states in \secr{genbsstr}.

\section{Physical meaning and realization of general boundary conditions \lbl{sec:realization}}

The general BCs \eqn{bc} have been formally derived from the fundamental principle of quantum mechanics, the norm conservation,
and there is a legitimate question about their physical meaning and what they represent for real systems. We address this question now.

Consider a ``microscopic'' model~\cite{micro} with a bulk Hamiltonian $\Hch(W_\x{bulk})$ characterized by the set $W_\x{bulk}$ of parameters
and a boundary characterized by the set $W_\x{boundary}$ of parameters.
Such microscopic model could be a lattice model, in which case the boundary is described by a specific termination,
or a different CM with more degrees of freedom (more wave-function components and/or higher momentum powers),
with the boundary described by some {\em specific} BCs. Assume that there is a well-defined low-energy limit of this microscopic model.
Let $\Hh(\ph;W_\x{bulk}^\x{low})$ be the continuum Hamiltonian to which $\Hch(W_\x{bulk})$ reduces in this limit.
Typically, among the parameters $W_\x{bulk}=(W_\x{bulk}^\x{low},W_\x{bulk}^\x{high})$
of the microscopic Hamiltonian $\Hch(W_\x{bulk})$ there are some $W_\x{bulk}^\x{high}$
that do not enter the low-energy Hamiltonian $\Hh(\ph;W_\x{bulk}^\x{low})$:
$W_\x{bulk}^\x{high}$ could be describing regions in the full Brillouin zone
away from the expansion points and/or higher momentum powers in the expansion in the microscopic CM.

Since the family of general BCs \eqn{bc} for the continuum bulk Hamiltonian $\Hh(\ph;W_\x{bulk}^\x{low})$ is exhaustive (includes all possible BCs),
the specific boundary described by $W_\x{boundary}$ in the microscopic model
will necessarily be described in the low-energy model by the BCs \eqn{bc} with a specific instance $\Uh(W_\x{bulk}^\x{high},W_\x{boundary})$
of the unitary matrix determined by the set of parameters $(W_\x{bulk}^\x{high},W_\x{boundary})$.
[In the microscopic model, the boundary must, of course, be defined in accord with the norm-conservation principle;
therefore, the latter will be automatically satisfied in the low-energy model as well, and its BCs are presentable in the form \eqn{bc}.]
This matrix $\Uh(W_\x{bulk}^\x{high},W_\x{boundary})$ will generally be determined by both the structure of the boundary,
described by the parameters $W_\x{boundary}$,
and the high-energy parameters $W_\x{bulk}^\x{high}$ of the microscopic bulk Hamiltonian $\Hch(W_\x{bulk})$.
This way, all necessary information about the microscopic structure of the boundary
and higher-energy part of the bulk spectrum of the microscopic model is contained in these BCs.
Looking at this relation in the reverse order, 
any microscopic model of a system with a boundary that has a given continuum model
with specific Hamiltonian $\Hh(\ph)$ and matrix $\Uh$ of the BCs as its low-energy limit,
can be regarded as a {\em (microscopic) realization} of the latter.

If the microscopic model $\{\Hch(W_\x{bulk}^\x{low},W_\x{bulk}^\x{high}),W_\x{boundary}\}$
allows for variation of the parameters $(W_\x{bulk}^\x{high},W_\x{boundary})~\in~\Wc$
within some domain $\Wc$, then the corresponding set of BCs with the matrices $\{\Uh(W_\x{bulk}^\x{high},W_\x{boundary}),
(W_\x{bulk}^\x{high},W_\x{boundary})\in\Wc\}$ can be realized
for a {\em fixed form} of the Hamiltonian $\Hh(\ph;W_\x{bulk}^\x{low})$ of the CM, specified by the low-energy parameters $W_\x{bulk}^\x{low}$.
For a given microscopic model with a moderate number of variable parameters,
it can be that {\em only a subset} of the whole parameter space $\Ux(\Nc/2)$ of possible BCs is spanned this way.
However, with several different microscopic models that are represented by the same low-energy Hamiltonian $\Hh(\ph;W_\x{bulk}^\x{low})$,
the whole BCs parameter space can always be spanned.
Alternatively, it should always be possible to construct a general enough microscopic model,
within which the whole BCs parameter space of the low-energy CM can be spanned.

Therefore, we conclude that the physical meaning of the family of general BCs \eqn{bc} satisfying the norm-conservation principle
is that it asymptotically describes {\em all possible} underlying microscopic models,
with all possible structures of their boundaries and all possible behaviors
of their bulk Hamiltonians in the Brillouin zone away from the expansion points,
that are represented by a given continuum Hamiltonian $\Hh(\ph)$ [\eq{H}] in the low-energy limit.
The implications of this for the bound states are discussed in the next section.

When the fully specified microscopic model $\{\Hch(W_\x{bulk}^\x{low},W_\x{bulk}^\x{high}), W_\x{boundary}\}$ of a system with a boundary is provided,
both the Hamiltonian $\Hh(\ph,W_\x{bulk}^\x{low})$ and the BCs, characterized by $\Uh(W_\x{bulk}^\x{high},W_\x{boundary})$,
of the CM can be derived from this microscopic model.
For Hamiltonians, such systematic low-energy-expansion procedure that consistently eliminates higher-energy degrees of freedom
is well-known~\cite{Winkler} and is part of the $k\cd p$ method.
It can also always be adapted to include the derivation of the corresponding BCs.
The main step is presenting the microscopic wave function as an expansion in terms of low-energy envelopes in the coordinate space,
but, for the derivation of the BCs, inclusion of the decaying short-spatial-scale particular solutions, if present, is also necessary.

Here we list a few examples of the models, for which such procedure of deriving the BCs has been performed.
There are variants of this procedure, depending on the type of the model,
but usually it reduces to one of the following two procedures (or their combination):
(i) Eliminating high-energy bands, thereby decreasing the size of the local (in momentum) Hilbert space.
In \secr{connection}, we derive the BC for the low-energy quadratic-in-momentum one-component model ($N=1,M=2$),
which describes the linear-in-momentum two-component model ($N=2,M=1$) in the vicinity of the minimum of the upper band.
Here, the total number $\Nc=2$ of degrees of freedom stays the same, but the wave-function components
are effectively ``converted'' into the power of momentum.
In Ref.~\ocite{KharitonovLSM}, BCs were derived for the quadratic-node Luttinger semimetal model~\cite{Luttinger1956}
from the Kane model with hard-wall BCs by eliminating the states with the angular momentum $\f12$.
This allowed us to demonstrate that the Luttinger semimetal exhibits one or two bands of surface states.
In Refs.~\cite{Samokhin,KharitonovSC}, BCs were derived for the linear-in-momentum model of the 1D superconductor
from the quadratic-in-momentum model with the hard-wall BCs.

(ii) Eliminating higher-order momentum terms in CMs within the same local Hilbert space.
In Ref.~\ocite{KharitonovQAH}, BCs were derived for the linear-in-momentum model of a quantum anomalous Hall system
from the quadratic-in-momentum model with hard-wall BCs (one block of the Bernevig-Hughes-Zhang model~\cite{BHZ})
in the vicinity of the topological phase transition.
A similar procedure arises when the microscopic model is a lattice model.
In Ref.~\ocite{AkhmerovPRB}, BCs for the linear-in-momentum model of graphene were derived from the lattice model.

Note that when instances of the BCs for the low-energy model are derived this way from the microscopic model,
they will be delivered in a completely arbitrary form;
the structure of the standardized form \eqn{bc} is not guaranteed to be identified at all.
Also, the current-nullification constraint is not applied explicitly as a derivation principle,
but rather will be automatically satisfied, since it must already be satisfied in the microscopic model.

Concluding this part, we point out that, analogously to the situation with the Hamiltonians in the $k\cd p$ method,
there are two ``sides'' 
of the formalism of general BCs that should not be conflated:
(i) derivation of the general BCs from general principles (norm conservation and, if present, symmetries, see the next \secr{schemes})
and exploring their whole parameter space;
(ii) derivation of instances of the general BCs from the underlying microscopic models.

\section{General boundary conditions deliver general bound-state structures\lbl{sec:genbsstr}}

The main value of the formalism of general BCs for CMs lies in what it provides for the study of bound states.
As follows from the previous \secr{realization},
since the general BCs \eqn{bc} describe all possible boundaries,
the CM with these BCs will capture {\em all possible} bound-state structures
of all microscopic models represented by the considered continuum Hamiltonian in the low-energy limit.
All properties of bound states that originate at low energies will be present in such CM.
The bound-state structures of the CM will asymptotically agree with those of the microscopic models from which the CM originates,
in the region of momenta and energies where the CM is valid.
According to the nonredundant one-to-one parametrization of the family of general BCs by unitary matrices (\secr{bc}),
one instance of $\Uh$ provides one instance of possible BCs, with one specific respective bound-state structure.
Spanning the parameter space $\Ux(\Nc/2)$ of $\Uh$, all possible bound-state structures for the CM Hamiltonian $\Hh(\ph)$ are obtained.
The evolution of bound-state structures as a function of $\Uh$ provides their exhaustive characterization.

The semi-analytical method formulated in \secr{method}
allows one to fully exploit the advantages of the formalism in the study of bound states.
Owing to the relatively small number $\Nc$ of relevant degrees of freedom (wave-function components $M$ and orders of momentum $N_m$)
sufficient to capture the essential behavior in the low-energy limit,
such models are usually still simple enough that
the whole parameter spaces of their Hamiltonian (see \secr{schemes}) and general BCs can be fully explored and
their general bound-state structures can be found in a tractable and explicit manner.
For the simplest models, the general bound-state structure can be found entirely analytically.
For more complicated model, one can still carry out this procedure ``semi-analytically'',
with minimal or modest computational resources.
The latter may be required for finding particular solutions \eqn{chipart} and solving the final equation \eqn{eqc} for the bound-state energy.

Another apparent advantage of the semi-analytical method is that
bound states can be found for a truly half-infinite system; as a result,
the low-energy features (such as the vicinity of the nodes of semimetals in higher dimensions) can be resolved with any desired accuracy.
This appealing property may be lacking in other methods,
such as the commonly employed finite-size numerical calculations, where resolution is limited by spatial quantization effects.

\section{Application of the formalism to families of bulk systems \lbl{sec:schemes}}

\begin{figure}
\includegraphics[width=.55\textwidth]{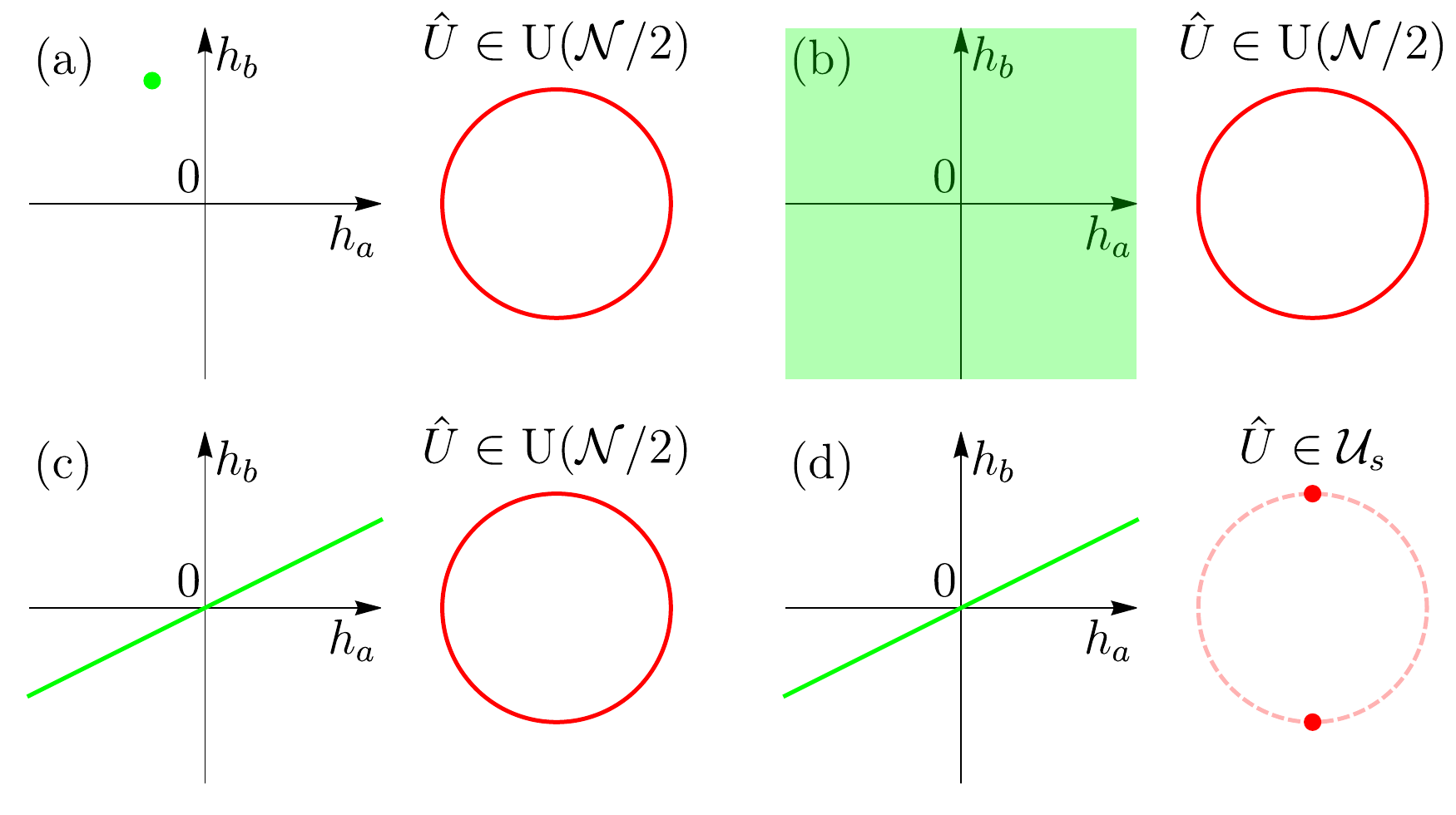}
\caption{
Schematic illustration of possible application schemes of the formalism of general BCs.
The space of parameters $h$ of the Hamiltonian $\Hh(\ph;h)$ is 
represented by a 2D plane $h=(h_a,h_b)$;
the parameter space $\Ux(\Nc/2)$ of the unitary matrices $\Uh$ of the general BCs \eqn{bc}
is 
represented by a circle (which is true for $\Nc=1$).
(a) The Hamiltonian $\Hh(\ph;h)$ is fixed ($h$ has a fixed value represented by a point),
the BCs are of the most general form, with $\Uh\in\Ux(\Nc/2)$ spanning the whole space.
(b), (c), (d) Applications schemes to {\em families} of bulk systems.
(b) Both the Hamiltonian $\Hh(\ph;h)$ and the BCs of are the most general forms,
with $h$ and $\Uh$ spanning their whole respective spaces.
(c) Symmetry-constrained family of the bulk Hamiltonians and general BCs.
The Hamiltonian $\Hh(\ph,h)$ with $h\in\Hc_s$ is of the most general form satisfying certain symmetries;
the symmetry-constrained parameter vector space $\Hc_s$ is represented by a line.
The BCs are of the most general form, with $\Uh\in\Ux(\Nc/2)$.
The model describes a family of bulk systems satisfying certain symmetries with all possible boundaries (which generally break the bulk symmetries).
(d) Symmetry-constrained families of both the bulk Hamiltonians and BCs.
The Hamiltonian $\Hh(\ph,h)$ with $h\in\Hc_s$ and BCs with $\Uh\in\Uc_s\subset \Ux(\Nc/2)$ are
of the most general forms satisfying certain symmetries;
the subset $\Uc_s$ is 
represented as consisting of two points, which is the case, e.g.,
for linear-in-momentum models with chiral or charge-conjugations symmetries~\cite{KharitonovLSM,KharitonovQAH,KharitonovFGCM}.
The model describes a family of systems with a boundary satisfying certain symmetries (a symmetry class).
The application scheme (d) is perfectly suited for the study of bound states in topological systems; 
the bound states within such models respect the symmetries and reflect bulk topology.
}
\lbl{fig:schemes}
\end{figure}

In this section, we outline possible application schemes of the formalism of general BCs for CMs.

\subsection{Symmetry-constrained family of Hamiltonians, general boundary conditions \lbl{sec:schemefamilyH}}

The main focus of the previous sections was the derivation of all possible BCs for a given Hamiltonian $\Hh(\ph)$, which was implied to be fixed.
According to \secr{genbsstr}, this scheme describes all possible boundaries and delivers an
exhaustive characterization of all possible bound-state structures for a {\em fixed} bulk system.
This application scheme is very useful, e.g., for a specific material, whose bulk structure is fixed, but different boundaries are possible.

However, even more powerful application schemes arise (\figr{schemes}) when general {\em families} of bulk systems are considered,
when, rather than a single fixed Hamiltonian $\Hh(\ph)$ [\figr{schemes}(a)],
a family of bulk Hamiltonians $\Hh(\ph;h)$, $h\in\Hc_s$, is considered, where the collection of parameters $h$ spans some space $\Hc_s$.
Oftentimes, these are families of Hamiltonians satisfying certain symmetries.
In this case, the most general form of Hamiltonians $\Hh(\ph;h)$ can be derived based solely on symmetries using the {\em method of invariants}:
for specified relevant degrees of freedom,
the wave functions components $\psi_m(x)$ (with their transformation properties) and order order $\Nc_m$ of momentum,
all invariant linearly independent matrix functions of momentum are found.
The most general form of the Hamiltonian $\Hh(\ph,h)$
is then an arbitrary linear combination of these basis functions and $h$ are the collection of its coefficients and $\Hc_s$ is their vector space.
This method is commonly used for ``conventional'' symmetries, such as spatial and time-reversal, and is known as the $k\cd p$ method.
However, it can also be readily applied to ``more abstract'' symmetries,
such as chiral and charge-conjugation, relevant to topological systems, see \secr{schemetopo}.

The general BCs parameterized by all possible unitary matrices $\Uh\in\Ux(\Nc/2)$
then represent {\em all possible boundaries} for the {\em whole family} of bulk systems described by the family of Hamiltonians $\Hh(\ph;h)$, $h\in\Hc_s$
[\figr{schemes}(c)],
Note that in the construction of the general BCs \eqn{bc},
only the current operator and the associated quantities, such as the vectors $\Psih_\pm[\psih(x);h]$ of right- and left-movers,
depend on the bulk parameters $h$, whereas the unitary matrices $\Uh$ specifying the BCs are independent parameters specifying possible boundaries.
Accordingly, this model will deliver the general bound-state structure for the whole family of bulk systems.

The scope of this application scheme is very wide since one family of CMs with a given symmetry
can describe many different real systems with many different lattice structures.
For example, crystals with different space groups but the same point group are described at the $\Ga$ point by one family of CMs.
The bulk crystal and its band structure can be quite complicated, and possible surfaces of such crystals -- even more so.
This approach allows one to obtain the general bound-state structures of such systems bypassing all these complications.

\subsection{Symmetry-constrained families of Hamiltonians and boundary conditions \lbl{sec:schemetopo}}

In the above application scheme for a symmetry-constrained family of bulk systems,
an arbitrary instance $\Uh\in\Ux(\Nc/2)$ of the BCs \eqn{bc} generally breaks the assumed symmetry of the bulk infinite system,
in which case the system {\em with a boundary} does not have that symmetry.
Symmetry of the system with a boundary is particularly important for topological systems and their bound states.

Topological systems are categorized according to the {\em symmetry classes},
with different sets of symmetries that protect various types of bulk topology of infinite systems.
Certain characteristics of the bound states of topological systems also rely on those symmetries
(such as the zero-energy bound states in 1D for particle-hole-type chiral or charge-conjugation symmetries;
counterpropagating edge states of $\Tc_-$-symmetric 2D topological insulators).
In order for the bound states to reflect (via bulk-boundary correspondence) the bulk topological properties
stemming from a certain symmetry, the system {\em with a boundary} must satisfy that symmetry.

Within the framework of CMs, the symmetry of the system with a boundary
means that not only the Hamiltonian, but also the BCs must also satisfy that symmetry~\cite{KharitonovLSM,KharitonovQAH,KharitonovSC,KharitonovFGCM}.
This requirement introduces constraints on the allowed form of $\Uh$.
As a result, the most general form of BCs under certain symmetries is parameterized by the matrices $\Uh\in\Uc_s$
that will typically belong to a subset $\Uc_s\subset\Ux(\Nc/2)$ of all possible unitary matrices [\figr{schemes}(d)].

Therefore, a CM obtained this way, in which both the bulk Hamiltonian and BCs are of the most general forms satisfying the symmetries,
represents a whole symmetry class of systems with a boundary.
As such, the model will deliver the general bound-state structure of the whole symmetry class.
This application scheme is therefore perfectly suited for the study of bound states in topological systems that rely on symmetries.

This application scheme has recently been formulated in Ref.~\ocite{KharitonovFGCM}
as a separate {\em (symmetry-incorporating) formalism of general continuum models with BCs}
and applied therein to the two-component linear-in-momentum model in 1D, 2D, and 3D.
Previously, it has also been applied~\cite{KharitonovSC} to 1D superconductors with the charge-conjugation symmetry $\Cc_+$, such that $\Cc_+^2=1$.

The general Hamiltonian $\Hh(\ph;h)$ [\eq{H}] with no symmetry constraints [\figr{schemes}(b)],
whose arbitrary matrices $\hh_n$ are seen as the free parameters $h$, supplemented with the general BCs \eqn{bc},
describes a family systems with the boundary of the symmetry class A, according to the topological classification.

\section{Quadratic-in-momentum ($N=2$) one-component ($M=1$) model\lbl{sec:N2M1}}

\subsection{Hamiltonian and general boundary condition\lbl{sec:H2bc}}

In this and the next sections, we demonstrate the application of the formalism of general BCs \eqn{bc}
to the simplest cases with the minimal number $\Nc=2$ of degrees of freedom necessary to have a well-defined boundary:
the models with one $\Nc_+=1$ right- and one $\Nc_-=1$ left-moving wave. There is {\em one} BC, the most general for of which reads
\beq
	\Psir_+[\psih(0)]=U\Psir_-[\psih(0)].
\lbl{eq:bcNc2}
\eeq
The vectors $\Psir_\pm[\psih(x)]$ are scalars and the unitary matrix
\beq
	U=\ex^{-\ix\nu}\in\Ux(1), \spc \nu\in[0,2\pi)\sim S^1,
\lbl{eq:U1}
\eeq
is scalar and can be parameterized by the angle $\nu$ on a unit circle.
According to the interpretation (\secr{interpretation}) of the form \eqn{bc} of the general BCs, $\nu$ can be regarded as the phase-shift angle.

For $\Nc=2$, there are only two possible types of CMs in terms of the makeup \eqn{Nc} of their degrees of freedom:
$(N,M)=(2,1)$, which is a quadratic-in-momentum one-component model, and $(N,M)=(1,2)$, which is a linear-in-momentum two-component model.
In this section, we consider the quadratic-in-momentum Hamiltonian
\beq
	\Hh_2(\ph)=h_2\ph^2
\lbl{eq:H2}
\eeq
for the one-component wave function $\psi(x)$.
Considering an additional linear term $h_1\ph$ does not bring anything new, since it can be absorbed by changing the momentum origin.
This analysis is therefore valid for the quadratic-in-momentum Hamiltonian of the most general form.
We assume $h_2>0$; the case of $h_2<0$ is analogous.

The vector \eqn{Psi} reads
\[
	\Psih[\psi(x)]=\lt(\ba{c}\psi(x)\\\prh\psi(x)\ea\rt).
\]
The current matrix is given by \eq{J2}. It is diagonalized by the eigenvectors [\eq{S}]
\[
	\sh_\pm=\f1{\sq2}\lt(\ba{c} 1\\\pm 1\ea\rt)
\]
with the eigenvalues $\pm h_2/l$, respectively. The corresponding projections \eqsn{Psipm}{Psir} are
\beq
	\Psi_\pm[\psi(x)]=\sh_\pm^\dg\Psih[\psi(x)]=\f1{\sq2}[\psi(x)\pm\prh\psi(x)],
\lbl{eq:PsipmH2}
\eeq
\beq
	\Psir_\pm[\psi(x)]=\sq{\f{h_2}{l}} \Psi_\pm[\psi(x)].
\lbl{eq:Psir2}
\eeq
The family of general BCs \eqn{bcNc2} for the quadratic Hamiltonian \eqn{H2},
spelled out in terms of the wave function, therefore reads
\beq
	\sq{\f{h_2}{l}}[\psi(0)+l\ph\psi(0)]=\ex^{-\ix\nu}\sq{\f{h_2}{l}} [\psi(0)-l\ph\psi(0)].
\lbl{eq:bcH2univ}
\eeq
One can rewrite this BC in the equivalent form
\beq
	\psi(0)=l\cot\tf\nu2\pd_x\psi(0).
\lbl{eq:bcH2}
\eeq
All possible BCs are parameterized  by the angle $\nu\in[0,2\pi)$ on the unit circle $S^1$ [\eq{U1}];
each value of $\nu$ corresponds to a distinct BC.

As explained in \secr{justification}, the fictitious length scale $l$ is explicitly present in the BC \eqn{bcH2},
as is the case for any model with momentum order $N>1$ higher than linear.
To better illustrate the roles of the fictitious length scale $l$ and parameter $\nu$,
we also derive the general BC for the Hamiltonian \eqn{H2} with the simpler ``poor man's'' approach.
The general linear homogeneous relation between the wave function and its derivative at the boundary has the form
\beq
	\psi(0)=L\pd_x\psi(0),
\lbl{eq:bcH2L}
\eeq
with initially a {\em complex} parameter $L$ of the physical dimension of length.
Imposing the nullification constraint \eqn{j=0} on the current [\eq{j123}]
\[
	j[\psi(x)]=h_2\{\psi^*(x)\ph\psi(x)+[\ph\psi(x)]^*\psi(x)\}
\]
of the Hamiltonian \eqn{H2} constrains $L$ to {\em real} values.
The two forms \eqsn{bcH2}{bcH2L} are equivalent upon the identification
\beq
	L=l\cot\tf{\nu}2
\lbl{eq:Lphi}
\eeq
and differ only in parametrization: all possible forms of the BC will be spanned upon $L\in(-\iy,+\iy)$
spanning the whole real axis space and upon $\nu\in[0,2\pi)$ spanning the unit circle for a fixed $l$.

This also explicitly demonstrates the important point explained in \secr{l} that the fictitious length parameter $l$
is not an additional to $\Uh$ free parameter of the family of general BCs: 
$l$ can have any arbitrary, but {\em fixed} value, when spanning all possible BCs by varying $\Uh$.
For a different value $l'$ of the length scale, the BC remains the same so long as $L$ remains the same,
i.e., the angle $\nu'=\nu'(\nu,l,l')$, satisfying $L=l\cot\f{\nu}2=l'\cot\f{\nu'}2$
will deliver the same BC for $l'$ as the BC \eqn{bcH2} with $l$ and $\nu$.

\subsection{Bound state for the general boundary condition\lbl{sec:H2bs}}

\begin{figure}
\includegraphics[width=.50\textwidth]{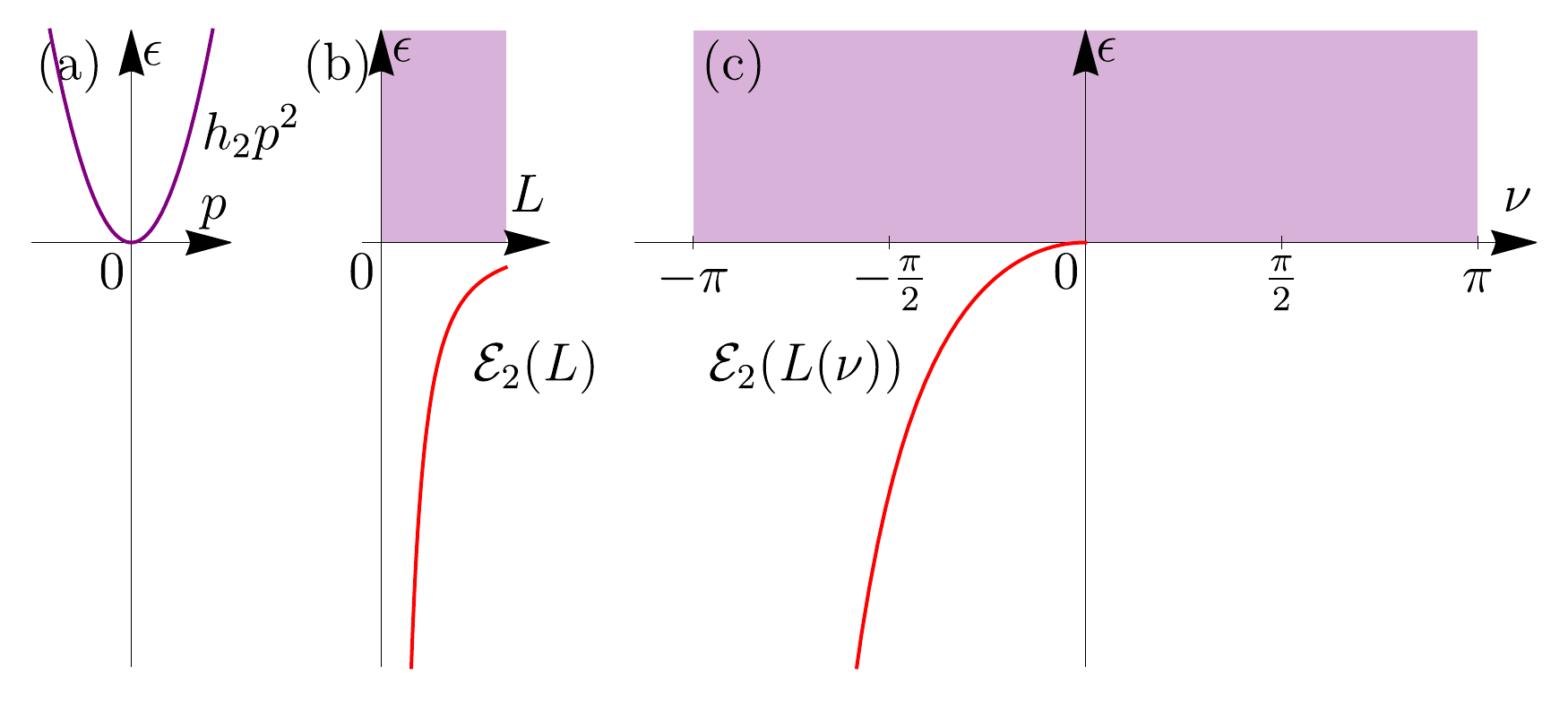}
\caption{(a) The bulk spectrum and (b) the bound state of the quadratic-in-momentum single-component model
with the Hamiltonian $\Hh_2(\ph)=h_2\ph^2$ [\eq{H2}] and general boundary condition \eqn{bcH2}.
A single bound state with the energy $\Ec_2(L(\nu))$ [\eqs{Lphi}{Ec2}, red] exists in the half $-\pi<\nu<0$
of the $\Ux(1)\sim S^1$ parameter space of the BC.
}
\lbl{fig:Ec2}
\end{figure}

For $h_2>0$, the bulk states fill the energy range $\e>0$ and bound states can exist only in the range $\e<0$.
We find that for $L<0$ [equivalently, for $\nu\in(-\pi,0)$ on the half-circle in \eq{bcH2}],
the model of a system occupying the half-line $x\geq0$ with the Hamiltonian $\Hh_2(\ph)$ [\eq{H2}] and BC \eqn{bcH2L}
contains one bound state with the wave function
\[
	\psi_\x{bs}(x)=\sq{\f{2}{|L|}}\ex^{-x/|L|}
\]
and energy
\beq
	\Ec_2(L)=-\f{h_2}{L^2},
\lbl{eq:Ec2}
\eeq
whereas for $L>0$ there are no bound states, \figr{Ec2}.
(This also shows that $L$ is perhaps a more convenient parameter than $\nu$ for the BC for this Hamiltonian.)

\subsection{Realizations of the boundary condition by potentials}

\begin{figure}
\includegraphics[width=.45\textwidth]{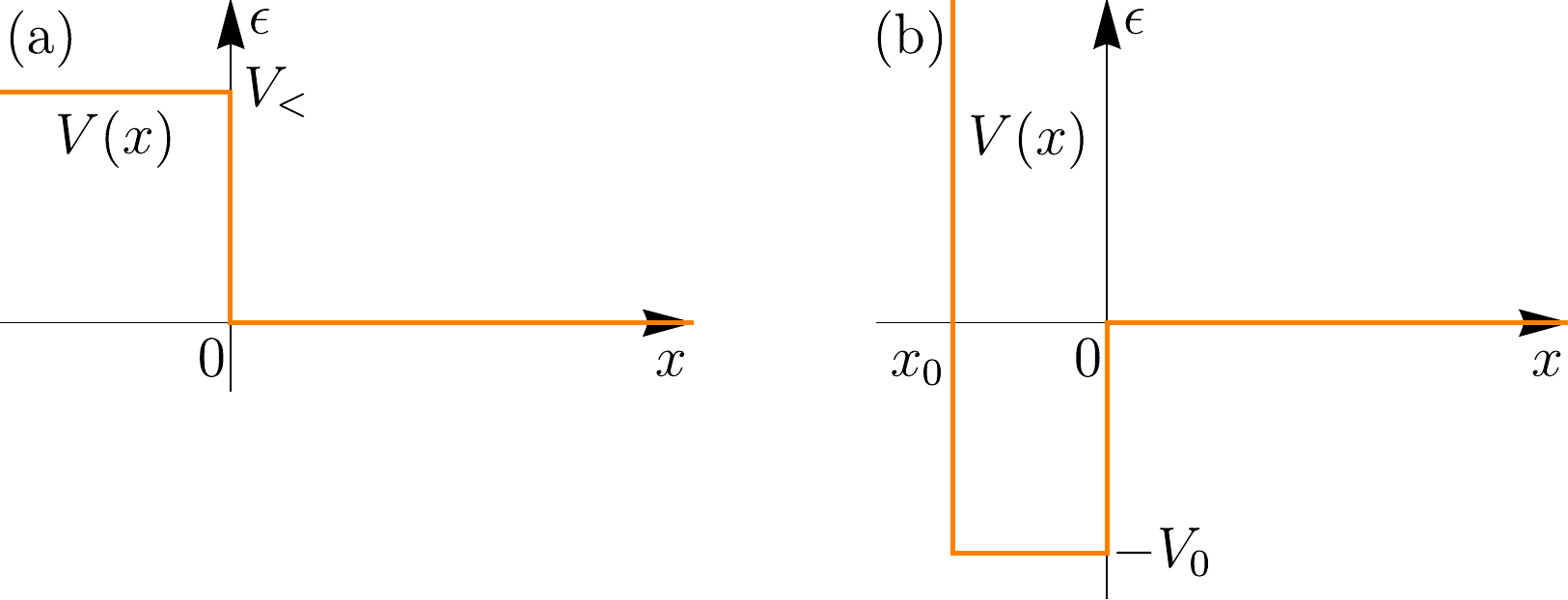}
\caption{
Realization of the general boundary condition \eqn{bcH2} by potentials
for the quadratic-in-momentum single-component model with the Hamiltonian $\Hh_2(\ph)=h_2\ph^2$.
(a) Potential wall \eqn{V1}. (b) Potential well \eqn{V2}.
}
\lbl{fig:V}
\end{figure}

We now consider one way the general BC \eqn{bcH2} for the Hamiltonian $\Hh_2(\ph)=h_2\ph^2$
can physically arise in the low-energy limit from a ``more microscopic'' model.
This is one example of the larger point of the {\em realization} of general BCs \eqn{bc}, discussed in \secr{realization}.
Another realization of the BC \eqn{bcH2} will be presented in \secr{connection}.

Consider the Hamiltonian
\beq
	\Hh_{2V}(\ph)=h_2\ph^2+V(x)
\lbl{eq:H2V}
\eeq
describing a Sch\"odinger particle in the potential $V(x)$ with the one-component wave function $\psi_V(x)$.

\subsubsection{Potential wall}

First, consider the potential ``wall''
\beq
	V(x)=\lt\{\ba{ll}
		0,& x>0,\\
		V_<>0,& x<0,\ea\rt.
\lbl{eq:V1}
\eeq
depicted in \figr{V}(a).
The wave function $\psi_V(x)$ is defined on the whole real axis $x\in(-\iy,+\iy)$; as such, there are no boundaries and no BCs.

The model \eqsn{H2V}{V1} is an example of a ``more microscopic'' model
related to the CM \eqsn{H2}{bcH2} of interest in the way described in \secr{system}:
in the region $x\geq0$, it has the latter as its low-energy limit, which in this case is simply the same,
and in the ``inaccessible'' region $x<0$ only ``high-energy'' degrees of freedom can exist, not captured by the low-energy model.
The BC for the low-energy wave function $\psi(x)$, defined only in the region $x\geq 0$,
arises upon systematically excluding these high-energy degrees of freedom. We demonstrate this procedure next.

Consider the stationary Schr\"odinger equation
\beq
	\Hh_{2V}(\ph)\psi_V(x)=\e\psi_V(x)
\lbl{eq:H2Veq}
\eeq
at energy $|\e|\ll V_<$.
At such low energies, the wave function $\psi_V(x)$ will rapidly decay into the region $x<0$
over a scale $\sq{h_2/V_<}$ much shorter than the typical length scale $\sq{h_2/|\e|}$ of its variation in the $x>0$ region.
The system becomes asymptotically equivalent to the one defined on the half-line $x\geq0$,
with the Hamiltonian $\Hh_2=h_2\ph^2$, the wave function $\psi(x)$, and an effective BC at $x=0$.
To obtain this BC, it is justified to completely neglect the energy $\e$ in the Schr\"odinger equation \eqn{H2Veq} in the region $x<0$,
which leads to the energy-independent solution
\[
	\psi_V(x)=B\ex^{\ka_<x},\spc x<0,
\]
with
$
	\ka_<=\sq{{V_<}/{h_2}}
$
and an arbitrary coefficient $B$.
The arbitrary wave function $\psi_V(x)$ at $x>0$ must match with this solution at $x=0$
according to the differential properties of the Hamiltonian $\Hh_{2V}(\ph)$ with the potential \eqn{V1}.
[Importantly, this implies that there is no ``more microscopic'' structure at the potential jump at $x=0$, 
i.e., that \eqs{H2V}{V1} is the ``fully microscopic'' Hamiltonian that determines the matching of the wave function.]
As follows from the stationary Schr\"odinger equation \eqn{H2Veq},
the wave function and its first derivative at $x=0$ on the right and left must be equal, which gives
\[
	B=\psi_V(+0),
\spc
	B\ka_<=\pd_x\psi_V(+0).
\]
Excluding the coefficient $B$ from these relations and equating $\psi_V(x)=\psi(x)$ in the region $x>0$, we obtain the effective BC
\beq
	\psi(0)=\f1{\ka_<}\pd_x\psi(0)
\lbl{eq:bc2wall}
\eeq
of the form \eqn{bcH2L} for the wave function $\psi(x)$, defined only at $x>0$.
Since $\ka_<>0$, there are no bound states in the low-energy model defined in the region $x>0$ with $\Hh_2=h_2\ph^2$ and BC \eqn{bc2wall},
in accord with the fact that there are no bound states in the exact solution for the Hamiltonian \eqn{H2V} with the potential \eqn{V1}.

Note also that, since the variation length scale $\sq{h_2/|\e|}$ of the wave function in the region $x>0$
is much larger than $1/\ka_<$ for $|\e|\ll V_<$, it is also justified to neglect the right hand-side in \eq{bc2wall} completely, and the BC reduces to
\beq
	\psi(0)=0.
\lbl{eq:hwbc}
\eeq
This is known as the {\em ``hard-wall'' BC}.
We observe that the hard-wall BC is realized asymptotically for a potential wall to the leading order in $|\e|/V_<\rarr+0$.

\subsubsection{Potential well \lbl{sec:potwell}}

Next, we consider the case of the potential well
\beq
	V(x)=\lt\{\ba{ll}
		0,&x>0,\\
		-V_0,& -x_0<x<0,\ea\rt.
\lbl{eq:V2}
\eeq
of width $x_0$ and depth $V_0>0$, depicted in \figr{V}(b).
To the left of the well, we consider an ``infinitely high'' potential wall;
according to above, the wave function $\psi_V(x)$ of the Hamiltonian \eqn{H2V} satisfies the hard-wall BC [\eq{hwbc}]
\beq
	\psi_V(-x_0)=0.
\lbl{eq:hwbc2}
\eeq
The wave function $\psi_V(x)$ of this system is defined in the half-infinite region $x\in(-x_0,+\iy)$.
Consider the energy $|\e|\ll V_0$ much smaller than the depth of the well.
Similarly, at such low energies, the system becomes asymptotically equivalent to the system defined only in the region $x>0$,
with the Hamiltonian $\Hh_2(\ph)=h_2\ph^2$, the wave function $\psi(x)$, and an effective BC for the latter at $x=0$.
Although the region $-x_0<x<0$ of the well is classically accessible, the excitation energy $\sim V_0$ there
is much higher than $\e$ and the region cannot be described by the low-energy wave function $\psi(x)$, see also the comment \ocite{gapcomment}.
One may neglect the energy $\e$ in the Schr\"odinger equation \eqn{H2Veq} in the region $-x_0<x<0$,
to obtain the energy-independent solution satisfying the BC \eqn{hwbc2},
\[
	\psi_V(x)=A\sin k_0(x+x_0), \spc -x_0<x<0,
\]
with $k_0=\sq{V_0/h_2}$ and an arbitrary coefficient $A$.
Analogously to the above, the wave function $\psi_V(x)$ in the region $x>0$ must match with this solution
according to the differential properties of the Hamiltonian $\Hh_{2V}(\ph)$ with the potential \eqn{V2}, which leads to the relations
\[
	A\sin k_0 x_0=\psi_V(+0),
\spc
	k_0 A\cos k_0 x_0=\pd_x\psi_V(+0).
\]
Excluding the coefficient $A$ and equating $\psi_V(x)=\psi(x)$ in the region $x>0$, we obtain the effective BC
\beq
	\psi(0)=L_0\pd_x\psi(0)
\lbl{eq:bc2well}
\eeq
of the form \eqn{bcH2L} with
\beq
	L_0=\f1{k_0}\f{\sin k_0 x_0}{\cos k_0 x_0}.
\lbl{eq:L0}
\eeq

Let us compare the properties of the microscopic [\eqss{H2V}{V2}{hwbc2}] and low-energy [$\Hh_2(\ph)=h_2\ph^2$, \eq{bc2well}] models.
On the one hand, according to \secr{H2bs}, for the low-energy model with $\Hh_2(\ph)=h_2\ph^2$ and general BC \eqn{bcH2},
there exists one bound state for $L<0$ with the energy \eqn{Ec2} and no bound state for $L>0$.
On the other hand, depending on the value of the parameter $k_0 x_0$, the microscopic model can have an arbitrary number of bound states,
whose energies $\e$ are determined from the equation
\beq
	\tan k_0(\e)x_0=-\f{k_0(\e)}{\ka(\e)},
	\spc k_0(\e)=\sq{\f{V_0+\e}{h_2}}, 
	\spc \ka(\e)=\sq{\f{-\e}{h_2}},
\lbl{eq:EcV2eq}
\eeq
obtained by solving the model exactly.
As $k_0x_0$ is increased (i.e., the well is made deeper and/or wider), new bound states emerge at $\e=0$ and move down in energy.
From this exact solution, the zeros $\al_n=\pi(n+\f12)$, $n=0,1,\ldots$, of $\cos k_0 x_0$ correspond precisely to the emergence of a new bound state;
exactly at such values of $k_0 x_0$ there is a non-normalizable solution with $\psi(x>0)=\x{const}$.
For $k_0x_0$ slightly larger than each zero $\al_n$,
there is a shallow bound state in addition to those already existing at $k_0x_0$ slightly below $\al_n$.
It is this shallow bound state that is captured and quantitatively accurately described in this limit $|k_0 x_0-\al_n|\ll 1$ by the low-energy model;
the asymptotic expansion of its exact energy obtained from \eq{EcV2eq} agrees with \eq{Ec2} with $L=L_0$ [\eq{L0}] in this limit.
When $k_0 x_0$ is away from zeros $\al_n$, the bound state is not shallow anymore
(its energy is outside of the validity range $|\e|\ll V_0$)
and is therefore no longer accurately described by the low-energy model.
Accordingly, away from zeros $\al_n$, the length scale $L_0\lesssim 1/k_0$ is much shorter
than the characteristic length scale $\sq{h_2/|\e|}$ of the low-energy model.
Only close to the zeros, $L_0$ becomes large, $k_0L_0\rarr \pm\iy$ as $k_0 x_0\rarr \al_n\pm 0$.
Note that this model illustrates that $\pd_x\psi(0)=0$ is legitimate case of the general BC \eqn{bcH2L}, with $L=\pm\iy$;
it describes, e.g., the microscopic system with a potential well at the boundary
in the borderline regime, when a bound state emerges right at zero energy.

\section{Linear-in-momentum ($N=1$) two-component ($M=2$) model \lbl{sec:N1M2}}

\subsection{Hamiltonian and general boundary condition \lbl{sec:H1bc}}

The second option of the model with the minimal number $\Nc=2$ of the degrees of freedom
is the model with the linear-in-momentum ($N=1$) Hamiltonian and the two-component ($M=2$) wave function.
Previously, general BCs for such model have been explored for specific forms of the Hamiltonian in
Refs.~\ocite{Berry,Ahari,Walter,Hashimoto2016,Hashimoto2019,KharitonovLSM,KharitonovQAH}.
The most general form of such model has been explored recently in Ref.~\ocite{KharitonovFGCM}.

Here, to illustrate the key properties of the formalism of general BCs,
we also consider only the specific, simplest form of the Hamiltonian:
\beq
	\Hh_1(\ph)=\tauh_z v_z \ph+\tauh_x d_x,
\lbl{eq:H1}
\eeq
which is also a special case of the general linear-in-momentum Hamiltonian \eq{H1gen} of \secr{interpretation}.
Here, $\tauh_{x,z}$ are the Pauli matrices and we assume $d_x>0$ and $v_z>0$.
The wave function is denoted as
\[
	\psih(x)=\lt(\ba{c} \psi_+(x) \\ \psi_-(x) \ea\rt).
\]

The current matrix \eqn{J1gen}
\beq
	\Jh=\tauh_z v_z
\lbl{eq:J1}
\eeq
and the current
\[
	j[\psih(x)]=\psih^\dg(x)\tauh_z v_z\psih(x)=v_z\psi_+^*(x)\psi_+(x)-v_z\psi_-^*(x)\psi_-(x)
\]
of $\Hh_1(\ph)$ are determined by the velocity matrix.
Since the velocity matrix is diagonal, the current is diagonalized by the wave-function components $\psi_\pm(x)$ themselves
(labeled according to the sign $\pm$ of their contribution to the current),
which have the character of the right- and left-moving waves ($\Nc_+=\Nc_-=1$), respectively.
The scalar projections \eqn{Psir} read
\beq
	\Psir_\pm[\psih(x)] =\sq{v_z} \psi_\pm(x)
\lbl{eq:Psir1}
\eeq
and the general BC \eqn{bcNc2}, spelled out in terms of the wave-function components, reads
\beq
	\sq{v_z}\psi_+(0)=\ex^{-\ix\nu}\sq{v_z}\psi_-(0).
\lbl{eq:bc1}
\eeq
As in \secr{H2bc}, all possible forms of the BC are parameterized by the phase-shift angle $\nu\in[0,2\pi)\sim S^1$;
each value of $\nu$ on the unit circle corresponds to a distinct BC.

\subsection{Bound state for the general boundary condition \lbl{sec:H1bs}}

\begin{figure}
\includegraphics[width=.60\textwidth]{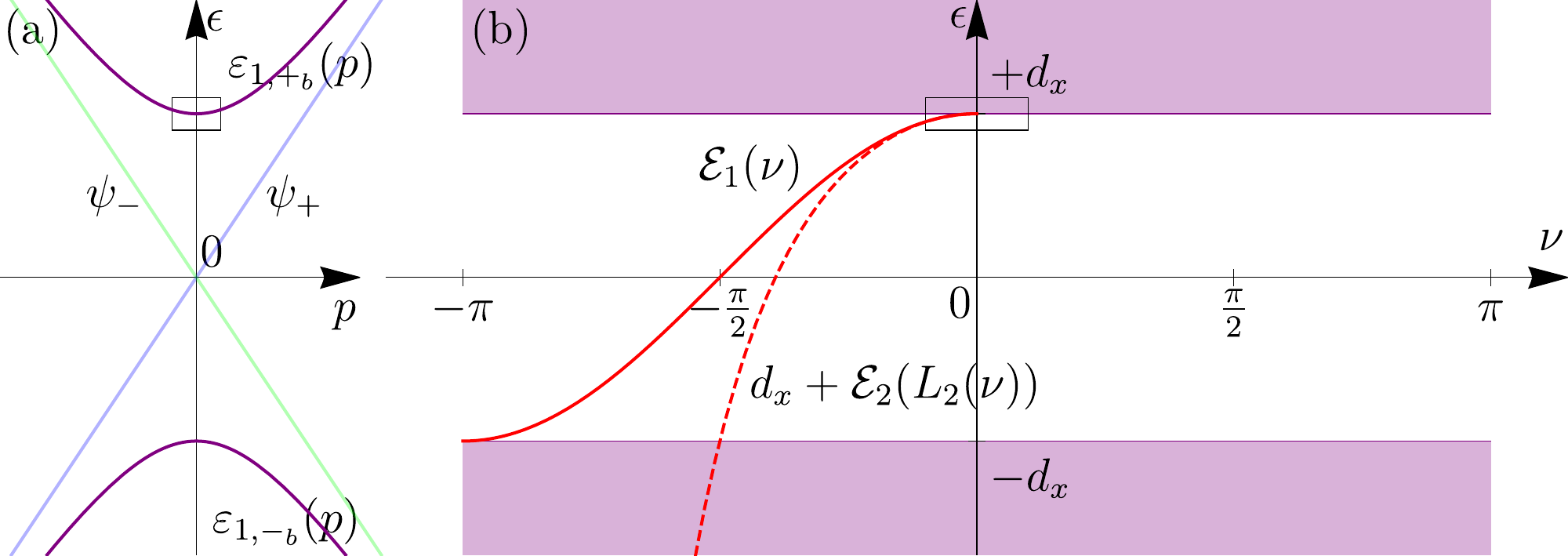}
\caption{
(a) The bulk spectrum and (b) the bound state of the linear-in-momentum two-component model with the Hamiltonian $\Hh_1(\ph)$ [\eq{H1}]
and general boundary condition \eqn{bc1}. A single bound state with the energy $\Ec_1(\nu)$ [\eq{Ec1}, red] exists in the half $-\pi<\nu<0$
of the $\Ux(1)\sim S^1$ parameter space of the BC.
The energy $d_x+\Ec_2(L_2(\nu))$ of the bound state of the quadratic-in-momentum one-component model [\eqs{H2}{bcH2}],
obtained as a low-energy expansion around the minimum of the upper band, is also shown in dashed red.
Their quantitative asymptotic agreement [\eqs{Ec1asymp}{Ec2asymp}] at small $\nu$ illustrates
the general connection (\secr{connection}) between the continuum models with the same total number $\Nc$ of degrees of freedom,
but different numbers of components $M$ and order of momentum $N$, $(M,N)=(1,2)$ and $(2,1)$ here.
The box schematically shows the validity range of the quadratic-in-momentum model, where the agreement is observed.
}
\lbl{fig:Ec1}
\end{figure}

The bulk spectrum of the Hamiltonian \eqn{H1} consists of two bands
\beq
	\eps_{1,\pm_b}(p)=\pm_b\sq{v_z^2p^2+d_x^2}.
\lbl{eq:e1}
\eeq
Bound states can exist only at energies in the gap region $\e\in(-d_x,d_x)$ between the bands.
Using the analytical method of \secr{method}, we find that, for the system occupying the half-line $x\geq0$,
there exists one bound state, with the wave function
\[
	\psih_\x{bs}(x)=\sq{\f{|p_\x{bs}|}{2}}\lt(\ba{c} \ex^{-\ix\f{\nu}2} \\ \ex^{+\ix\f{\nu}2} \ea\rt)\ex^{\ix p_\x{bs} x}
	,\spc
	p_\x{bs}=-\f{d_x}{v_z}\ix\sin\nu,
\]
and energy
\beq
	\Ec_1(\nu)=d_x\cos\nu,
\lbl{eq:Ec1}
\eeq
shown in \figr{Ec1}.
The bound state exists only in the range $-\pi<\nu<0$ of the phase-shift angle parameterizing the general BC \eqn{bc1},
i.e., in the half of its parameter space $\Ux(1)\sim S^1$.
The explanation and multiple general implications of this result have been provided in Ref.~\ocite{KharitonovFGCM}.

\section{Connection between the number of components and order of momentum\lbl{sec:connection}}

The same universal form \eqn{bc} of the general BCs for Hamiltonians with different
numbers of wave-function components $M$ and momentum orders $N_m$, 
but the same total number $\Nc=\sum_{m=1}^M N_m$ [\eq{Nc}]
of degrees of freedom, strongly suggests a connection between these two types of degrees of freedom.
Such connection can indeed be established and is quite insightful.
We demonstrate this connection for the two models considered above in \secsr{N2M1}{N1M2},
with the same minimal total number $\Nc=2$, but different composition in terms of $M$ and $N_m=N$.

The bulk bands \eqn{e1} of the linear-in-momentum ($N=1$) Hamiltonian \eqn{H1} for the two-component ($M=2$) wave function (\secr{N1M2})
are asymptotically quadratic in the vicinity of their extrema at $p=0$ (\figr{Ec1}):
\[
	\eps_{1,\pm_b}(p)=\pm_b\lt[d_x+\f{v_z^2}{2d_x} p^2+\Oc(p^4)\rt].
\]
For each of the bands, the system can asymptotically be described by a quadratic-in-momentum ($N=2$) Hamiltonian \eqn{H2}
for a one-component ($M=1$) wave function (\secr{N2M1}).

As discussed in more detail in \secr{realization},
the corresponding effective low-energy model can be derived using a systematic low-energy-expansion procedure.
Crucially, not only the bulk Hamiltonian (for which the procedure is well-known as part of the $k\cd p$ method~\cite{Winkler}),
but also the associated BC can always be derived within the same procedure.
We now present this procedure, which establishes the connection between the two models.

We do so for the minimum of the upper band $\eps_{1,+_b}(p)$. We consider the wave-function vector
\beq
	\psih(x)
	=\chih_{+_b} \psi_2(x)+\chih_{-_b} \de\psi(x)
\lbl{eq:psihrel}
\eeq
with components $\psi_2(x)$ and $\de\psi(x)$ in the basis of the eigenstates
\beq
	\chih_{\pm_b}=\f1{\sq2}\lt(\ba{c}1\\ \pm_b 1\ea\rt)
\lbl{eq:chix}
\eeq
at $p=0$, with the energies $\eps_{1,\pm_b}(0)=\pm_b d_x$, respectively.
The stationary Schr\"odinger equation in this basis reads
\[
	\lt[\lt(\ba{cc} d_x & v_z \ph \\ v_z\ph & -d_x\ea\rt)-\e\um\rt]\lt(\ba{c} \psi_2(z) \\ \de\psi(z) \ea\rt)=\nm,
\]
from which we express
\beq
	\de\psi(x)=\f1{\e+d_x}v_z\ph\psi_2(x)
\lbl{eq:dpsi}
\eeq
in terms of $\psi_2(x)$.
Inserting this form into the other equation, we obtain the closed equation
\beq
	d_x\psi_2+v_z\ph\f1{\e+d_x}v_z\ph\psi_2=\e\psi_2
\lbl{eq:psi2eq}
\eeq
for $\psi_2(x)$. At energies
\beq
	\e=d_x+\de\e
\lbl{eq:de}
\eeq
close to the upper-band minimum $\eps_{1,+}(0)=d_x$, such that $|\de\e|\ll d_x$, 
it is justified, to leading order, to substitute $\e\rarr d_x$ in the denominators in \eqs{dpsi}{psi2eq} to obtain
\beq
	\de\psi(x)=\f{v_z}{2d_x}\ph\psi_2,
\lbl{eq:dpsiapprox}
\eeq
\beq
	\f{v_z^2}{2d_x}\ph^2\psi_2(x)=\de\e\psi_2(x),
\lbl{eq:psih2eqapprox}
\eeq
respectively.
Equation~\eqn{psih2eqapprox} has the form of the stationary Schr\"odinger equation
\[
	\Hh_2(\ph)\psi_2(x)=\de\e\psi_2(x)
\]
for the wave function $\psi_2(x)$ with the Hamiltonian
\beq
	\Hh_2(\ph)=h_2\ph^2, \spc h_2=\f{v_z^2}{2d_x},
\lbl{eq:H2rel}
\eeq
and energy $\de\e$.
Equation~\eqn{dpsiapprox} provides the expression for the ``remainder'' $\de\psi(x)$.
For energies $\e$ in the vicinity of $d_x$ and momentum $p$ in the vicinity of $0$,
this remainder is parametrically small, $\de\psi(x)\ll \psi_2(x)$,
and the wave function \eqn{psihrel} has the dominant character of the $\chih_{+_b}$ eigenstate at $p=0$.
Nonetheless, this remainder is essential for the correct derivation of both the Hamiltonian \eqn{H2rel} above and the BC for it, as we now show.

To this accuracy, the two-component wave function \eqn{psihrel} of the linear Hamiltonian \eqn{H1} has the form [\eqs{chix}{dpsiapprox}]
\[
	\psih(x)=\f1{\sq2}\lt(\ba{c} \psi_2(x)+l_2\ph\psi_2(x) \\ \psi_2(x)-l_2\ph\psih_2(x)\ea\rt),
\]
with
\beq
	l_2=\f{v_z}{2d_x}.
\lbl{eq:l2rel}
\eeq
Inserting it into the BC \eqn{bc1}, we obtain the BC for the wave function $\psi_2(x)$ in the form
\beq
	\psi_2(0)+l_2\ph\psi_2(0)=\ex^{-\ix\nu}[\psi_2(0)-l_2\ph\psi_2(0)].
\lbl{eq:bcH2univ2}
\eeq

We recognize that this BC does have the form \eqn{bcH2univ} of the general BC for the quadratic-in-momentum Hamiltonian $\Hh_2(\ph)$ [\eq{H2}].
However, importantly, its parameters $\nu$ and $l_2$ are {\em not arbitrary},
but are determined by those of the initial linear-in-momentum two-component model \eqsn{H1}{bc1},
and so is the parameter $h_2$ of the bulk Hamiltonian \eqn{H2rel}:
the angle $\nu$ in \eq{bcH2univ2} is just that of the BC \eqn{bc1} and
the length scale $l_2$ has a specific value \eqn{l2rel} arising from the parameters of the Hamiltonian \eqn{H1}.
This is in contrast to the situation with the general BC \eqn{bcH2} for the quadratic model, where the angle $\nu$ and length scale $l$ are arbitrary.
This difference constitutes two conceptually different ``sides'' of the formalism of general BCs
and the relation between the two models is another example of the realization 
of the general BCs within a microscopic model, both general points discussed in \secr{realization}.

Further, we recognize that the scale $l_2$ is the ``microscopic'' scale
determining the {\em range of validity} ($p l_2\ll 1$, $|\de\e|\ll d_x$) of the low-energy model \eqsn{H2rel}{bcH2univ2}:
its wave function $\psi_2(x)$ must vary at scales much larger than $l_2$.

According to \secsr{N2M1}{N1M2}, for $-\pi<\nu<0$, both the initial linear-in-momentum model and its quadratic-in-momentum low-energy model
have a bound state, with the respective energies [\eqs{Ec2}{de}]
\beq
	\Ec_1(\nu)=d_x\cos\nu=d_x[1-\tf12\nu^2+\Oc(\nu^4)],
\lbl{eq:Ec1asymp}
\eeq
\beq
	d_x+\Ec_2(L_2(\nu))=d_x-\f{h_2}{L_2^2(\nu)}=d_x[1-\tf12\nu^2+\Oc(\nu^4)],
\spc
	L_2(\nu)=l_2\cot\tf\nu2.
\lbl{eq:Ec2asymp}
\eeq
We do observe quantitative asymptotic agreement between these energies of the bound state to the leading order in $|\nu|\ll1$, as shown in \figr{Ec1}.
For larger $|\nu|\sim 1$, outside of the validity range $|\de\e|\ll d_x$ of the quadratic-in-momentum low-energy model, the energies differ.

This establishes the relation between both the Hamiltonians and BCs of the models with same total number $\Nc=2$ of degrees of freedom,
but different numbers of the wave-function components $M$ and orders of momentum $N$.
We see that this relation is not a one-to-one correspondence.
Rather, the system with less components and higher momentum order ($N=2$, $M=1$)
is the low-energy limit of the system with more components and lower momentum order ($N=1$, $M=2$)
and therefore, the validity range of the former is smaller.
Analogous relations can be established between models with more degrees of freedom, $\Nc\geq 4$.

\section{Generalizations and extensions of the formalism to higher dimensions, junctions, and interfaces \lbl{sec:generalizations}}

The formalism of general BCs, presented in this work for a half-infinite 1D system,
allows for a number of natural generalizations and extensions to systems with other geometries.
The fundamental principle of the wave-function-norm conservation always holds,
and to derive general BCs, a similar procedure of finding all possible BCs
that satisfy the current conservation principle has to be carried out.

First, the formalism could be extended to systems of higher dimensions.
One can immediately distinguish between the cases of preserved and broken translation symmetry along the surface.
For preserved translation symmetry,
the momentum along the surface is conserved and becomes a parameter for the family of effective 1D Hamiltonian
for the motion perpendicular to the surface.
Hence, one could expect the problem to reduce to the 1D problem (possibly, with some necessary adaptations).
When translation symmetry along the surface is broken (due to, e.g., surface roughness, disorder), e.g., as considered in Ref.~\ocite{Walter},
the problem becomes more complicated and may not readily reduce to an effective 1D problem.

Next, one could adapt this formalism to junctions and interfaces~\cite{Tokatly} between the systems described by different CMs,
which could have completely different band structures.
The current continuity at the junction or interface is still the guiding principle that leads to the general BCs.
Equivalently, one can reduce the geometry
to that of one half-infinite system by mapping all half-infinite subsystems onto one region $x>0$.
For example, in the case of two coupled 1D systems, one at $x>0$ and the other at $x<0$,
with the respective currents $j_\gtrless(x)$, one can map the $x<0$ system onto the $x>0$ region by the change of coordinate $x\rarr-x$,
in which case the net current of the new system will be $j(x)=j_>(x)-j_<(-x)$.
The current continuity requirement $j_<(0)=j_>(0)$ is then equivalent to the current nullification requirement $j(0)=0$.

The application schemes outlined above in \secr{schemes} should readily apply to these other geometries as well, once the formalism has been generalized.

\section{Conclusion and outlook \lbl{sec:conclusion}}

In this work, we deliberately focused only on the derivation, substantiation, elucidation, and expansion
of the formalism of general BCs, while leaving its numerous possible applications to future works.
Already the existing
applications~\cite{AkhiezerGlazman,ReedSimon,Berry,BerezinShubin,Bonneau,Tokatly,McCann,AkhmerovPRL,AkhmerovPRB,Ostaay,Hashimoto2016,Hashimoto2019,
Ahari,KharitonovLSM,KharitonovQAH,Seradjeh,Walter,Shtanko,KharitonovSC,Enaldiev2015,Volkov2016,Devizorova2017,KharitonovFGCM}
demonstrate the advantages of the formalism, summarized in \secr{genbsstr}.
In all application scenarios, the formalism provides an exhaustive characterization of the bound-state structure.
Therefore, perhaps most importantly,
the formalism of general BCs allows one to not only check and confirm the expected properties of bounds states (such as topological properties),
but also to discover new effects and features, which may turn out to be quite unusual and unanticipated.

\begin{acknowledgments}
The author is thankful to B. Trauzettel, E. M. Hankiewicz, V. N. Golovach, and F. S. Bergeret for useful discussions
and to Deutsche Forschungsgemeinschaft (DFG) for financial support through the Grant No. KH 461/1-1.
\end{acknowledgments}

\appendix
\section{Possible redundancies of the parametrization of general boundary conditions \lbl{app:nonredundance}}

One of the key values of the universal form \eqn{bc} of general BCs
is that they are parameterized by unitary matrices in a nonredundant one-to-one (bijective) way:
each unitary matrix specifies one set of BCs corresponding to one admissible Hilbert space.

Since one can derive the general BCs by resolving the current-nullification constraint in many different ways,
it is worthwhile to point out the redundancies that may arise when the procedure is performed arbitrarily.
Two types of redundancies can be pointed out.

(i) Equivalence transformations. One can perform equivalence transformations upon the BCs.
These do not change the Hilbert space and are therefore redundant.
It is important to make sure that the parameters of the equivalence transformations are not regarded as meaningful parameters of the BCs.

(ii) Additional linearly dependent relations, i.e.,
when there are more than $\La=\Nc/2$ relations of which only $\La$ are linearly independent.
As an example, for linear-in-momentum Hamiltonians ($N=1$) for the multicomponent ($M$) wave function $\psih(x)$,
the BCs are often presented in the form~\cite{McCann,AkhmerovPRL,AkhmerovPRB,Shtanko,Hashimoto2016,Hashimoto2019}
\[
	\Mh\psih(0)=\nm,
\]
where $\Mh$ is a square matrix of dimensions $M\tm M$.
Since in the correct BCs there can only be $\La=\Nc/2=M/2$ linearly independent relations,
the rank of the matrix $\Mh$ is only $M/2$ and half of the relations in this form are redundant:
one could keep only $M/2$ linearly independent relations, which can always be presented in the form \eqn{bc}.

\section{$\Jh$ for $N=4$ \lbl{app:JN4}}

As an explicit illustration of the general property of the eigenvalues of the current matrix presented in \secr{justification},
consider the fourth-order-in-momentum ($N=4$) Hamiltonian $\Hh(\ph)=h_2\ph^2+h_4 \ph^4$
for a one-component ($M=1$) wave function. Its current matrix \eqn{J} reads
\[
	\Jh=\lt(\ba{cccc}
		0&J_2&0&J_4\\
		J_2&0&J_4&0\\
		0&J_4&0&0\\
		J_4&0&0&0\\
		\ea\rt),
\spc
	J_2=\f{h_2}l,\spc J_4=\f{h_4}{l^3}.
\]
The eigenvalues of $\Jh$ are
\[
	\Jc_{\pm_a\pm_b}(h_2,h_4;l)
	=\f12\lt(\pm_a\sq{4J_4^2+J_2^2}\pm_b J_2\rt)
	=\f12\lt[\pm_a\sq{4\lt(\f{h_4}{l^3}\rt)^2+\lt(\f{h_2}l\rt)^2}\pm_b \lt(\f{h_2}l\rt)^2\rt].
\]
We see that, indeed, there are two positive $\Jc_{+_a\pm_b}>0$ and two negative $\Jc_{-_a\pm_b}<0$ eigenvalues
and their signs remain the same regardless of the relation $h_4/(h_2 l^2)$ between the parameters;
in particular, they are the same as in the limit of dominant $h_4$, when $\Jc_{\pm_a\pm_b}(|h_4|\gg |h_2| l^2)=\pm_a|h_4|/l^3$.


\begin{thebibliography}{99}
\bibitem{gapcomment}
Although a ``gapped'', i.e., classically forbidden, region would be most common microscopic realization of the inaccessible region,
the latter could also contain a classically allowed region, so long as it is a region of ``high-energy motion''
not affected by the low energy scale of the CM.
An example is the case of the deep and narrow quantum well presented in \secr{potwell}.
\bibitem{micro}
We will refer to these models  as ``microscopic'' in relation to the considered continuum model
in the sense that they contains more degrees of freedom or the structure of their boundary is specified.
These models could, however, themselves be continuum models with some specific BCs
and arise as low-energy models from other, ``more microscopic'' models.
\end{thebibliography}
\end{document}